\documentclass[lettersize,journal]{IEEEtran}
\usepackage{amsmath,amsfonts}
\usepackage{algorithmic}
\usepackage{algorithm}
\usepackage{array}
\usepackage[caption=false,font=normalsize,labelfont=sf,textfont=sf]{subfig}
\usepackage{textcomp}
\usepackage{stfloats}
\usepackage{url}
\usepackage{verbatim}
\usepackage{graphicx}
\usepackage{cite}
\usepackage{enumitem}
\usepackage{booktabs}
\usepackage{multirow}
\usepackage{makecell}
\usepackage{xcolor,soul}

\begin{document}

\title{Event-triggered Implicit Perturbation for Zeroth-Order Fine-Tuning of Spiking Transformers}

\author{Tengteng Lei, Prabodh Katti, Rashi Dutt, Houssem Sifaou, Tan Peng, \\ Osvaldo Simeone, Kai Xu, Bipin Rajendran
        % <-this % stops a space
\thanks{T. Lei, P. Katti, R. Dutt, H. Sifaou, O. Simeone, and B. Rajendran are affiliated with the Institute for Intelligent Networked Systems (INSI), Northeastern University London, London E1 8PH, U.K. T. Peng and K. Xu are with the Centre for Intelligent Information Processing Systems (CIIPS), Department of Engineering, King’s College London, London WC2R 2LS, U.K. (b.rajendran@nulondon.ac.uk).

This project is funded by the Advanced Research + Invention Agency (ARIA). The work of O. Simeone and B. Rajendran was also supported by Open Fellowships of the EPSRC (EP/W024101/1,  EP/X011356/1) and by the EPSRC project (EP/X011852/1). The work of O. Simeone was also supported by  the European Research Council (ERC) under the European Union’s Horizon Europe Programme (grant agreement No. 101198347).}} % <-this % stops a space

% \markboth{IEEE Transactions,~Vol.~14, No.~8, August~2021}%
% {Shell \MakeLowercase{\textit{T. Lei et al.}}: A Sample Article Using IEEEtran.cls for IEEE Journals}


\maketitle

\begin{abstract}
Zeroth-order (ZO) optimization estimates gradients using only forward-pass evaluations, making it suitable for fine-tuning non-differentiable, event-driven spiking neural networks (SNNs). However, its deployment on in-memory computing (IMC) accelerators is constrained by the repeated read-modify-write (RMW) operations arising from explicit weight perturbation and the prohibitive hardware footprint of random number generators (RNGs) for statistically independent per-weight perturbations. To address these challenges, we propose an implicit-perturbation ZO (IPZO) architecture in which perturbation sums computed by an event-triggered perturbation generation unit (PGU) are combined with the weighted sums produced by the IMC array, eliminating perturbation-induced RMW operations while preserving weight-stationary execution of IMC. By exploiting spike sparsity, the PGU generates and accumulates perturbation contributions only for spike-activated weight rows, reducing the required row dimension of the RNG array. An address-driven XOR recombination scheme (PGU-XOR) is further introduced to mitigate the spatial correlations caused by direct RNG reuse (PGU-Reuse). The results show that (1) PGU-XOR matches software RNGs in accuracy on Spikingformer/CIFAR-10 ($76.41$\% vs. $76.53$\%) and perplexity (PPL) on SpikeGPT/WikiText-2 ($54.20$ vs. $53.23$), whereas PGU-Reuse degrades accuracy by $9.56$ percentage points and increases PPL by $11.8$; (2) implemented in a TSMC 16-nm CMOS technology, PGU-XOR incurs $40.3$\%--$46.0$\% area overhead and $15.2$\%--$48.9$\% energy overhead per matrix-vector multiplication (MVM) relative to PGU-Reuse, yet its faster convergence reduces the total perturbation energy to $0.51\times$ that of PGU-Reuse at iso-accuracy; (3) IPZO reduces the perturbation energy to $0.46\times$--$0.83\times$ that of conventional explicit weight perturbation for a batch size of $B=64$ and $T=4$ time steps, with the advantage growing as $BT$ decreases. 
\end{abstract}


\section{Introduction}
\IEEEPARstart{T}{he} transformer architecture has established itself as a dominant framework in artificial intelligence, driving state-of-the-art advances across fields ranging from natural language processing \cite{vaswani2017attention, wolf2020transformers} to computer vision \cite{wu2020visual, han2022survey}. Despite this success, the intensive computation and large memory footprint of transformers often preclude their efficient deployment on resource-constrained edge platforms, a challenge further exacerbated by the memory-wall bottleneck of traditional Von Neumann architectures \cite{wolters2024memory}. To mitigate these hardware limitations, neuromorphic accelerators integrating spiking transformers with in-memory computing (IMC) have emerged as a promising paradigm \cite{song2025xpikeformer, das2025aster, jiang2025sparta}. By executing matrix-vector multiplications (MVMs) directly within memory arrays, IMC eliminates prohibitive data-movement overhead between memory and computation. Binary spike activations further simplify MVM computation by reducing multiply-accumulate (MAC) operations to additions. Spike sparsity, in turn, limits synaptic computation to active events, thereby reducing unnecessary operations and energy consumption \cite{roy2019towards, davies2018loihi}. While existing accelerators are highly optimized for inference \cite{keller202395, mun2026asap, nag2023vita}, achieving efficient on-chip learning for adaptive and privacy-preserving edge intelligence remains a formidable challenge. Compared with inference, training incurs substantially higher memory and computational overhead, placing stringent constraints on hardware implementation \cite{wu202499, wang202422nm, shen2024efficient}.  

Zeroth-order (ZO) optimization has attracted growing attention for on-chip learning as a memory-efficient alternative to conventional backpropagation (BP) \cite{zhang2024revisiting, sugiura2025elasticzo, gu2021efficient}. Unlike BP, which requires storing intermediate activations and computing gradients through a backward pass \cite{han2023chip, dampfhoffer2023backpropagation}, ZO optimization estimates gradients using only forward-pass evaluations \cite{liu2020primer, larson2019derivative}. This distinction is particularly important for spiking neural networks (SNNs), where backpropagation through time (BPTT) requires retaining neuron states across multiple time steps, further increasing the memory overhead of training \cite{meng2023towards, zhang2024memory}. The forward-only nature of ZO optimization is naturally compatible with non-differentiable activations, making it well suited for training event-driven SNNs on resource-constrained edge hardware.

However, efficiently implementing ZO optimization in IMC-based accelerators is far from straightforward due to the mismatch between its algorithmic requirements and the underlying hardware architecture. Firstly, ZO optimization requires statistically independent random perturbations for every weight parameter. Although uniformly distributed perturbations have been demonstrated as an effective alternative to Gaussian perturbations \cite{tan2025perturbation, qinzeroth, liu2018zeroth}, supporting per-weight generation of uniform random numbers still incurs a prohibitive hardware cost. Even when implemented using compact linear-feedback shift registers (LFSRs), the area footprint of a single-bit uniform random number generator (RNG) far exceeds that of a memory cell storing one weight bit in an IMC array. This footprint mismatch limits the scalability of per-weight perturbation generation in IMC-based accelerators. While random-number reuse can alleviate the RNG overhead \cite{tan2025perturbation}, it introduces spatial correlations among the resulting perturbations and degrades the learning accuracy with insufficient random numbers. Secondly, conventional explicit-perturbation ZO optimization (EPZO) requires repeatedly reading the entire weight matrix, applying the perturbations, and writing the perturbed weights back into the IMC array during each forward pass for gradient estimation. The resulting frequent read-modify-write (RMW) operations compromise the energy advantage of IMC by reintroducing extensive data movement  \cite{sinangil20207, jiang2020c3sram, yeh202616nm}.

To overcome these bottlenecks, we propose an implicit-perturbation ZO optimization (IPZO) architecture for IMC-based SNN accelerators that combines event-triggered perturbation generation with accumulation-domain perturbation injection. Specifically, an event-triggered perturbation generation unit (PGU) generates and accumulates perturbation contributions only for spike-activated weight rows at each time step. Given the inherent sparsity of SNNs, the PGU thus reduces the required row dimension of the RNG array well below that of the weight matrix. To mitigate the spatial correlations introduced by directly reusing perturbations from the reduced RNG array across weight rows, an address-driven XOR recombination scheme is employed to construct independent perturbations across the entire weight matrix. Furthermore, accumulation-domain perturbation injection combines the perturbation sums computed by the PGU with the weighted sums produced by the IMC array. By keeping the weights stationary throughout each optimization step, IPZO eliminates repeated perturbation-induced RMW operations while preserving the energy efficiency of IMC.

The effectiveness of IPZO is demonstrated through comprehensive evaluations spanning algorithmic validation and hardware implementation, highlighting its potential for efficient on-chip learning in IMC-based SNN accelerators. The main contributions of this work are summarized as follows:

\begin{itemize}

\item We propose an IPZO architecture that injects perturbation contributions into the accumulation domain rather than the stored weight domain, preserving weight-stationary execution and eliminating the repeated RMW operations required by EPZO.

\item We develop an event-triggered perturbation generation unit (PGU) that generates perturbations only for spike-activated weight rows, decoupling the RNG array size from the weight matrix dimensions. To eliminate the spatial correlations introduced by direct RNG reuse, we design an address-driven XOR recombination scheme within the PGU.

\item We validate the proposed framework through algorithmic evaluation and post-layout hardware implementation in TSMC 16-nm CMOS. PGU-XOR achieves performance comparable to software RNGs across SNN image classification and language modeling, while its faster convergence reduces the total perturbation energy to $0.51\times$ that of PGU-Reuse. At the architecture level, IPZO equipped with PGU-XOR reduces perturbation energy to $0.46$--$0.83\times$ that of EPZO under practical SNN fine-tuning settings.

\end{itemize}

The remainder of this paper is organized as follows. Section II introduces the necessary preliminaries, and Section III analyzes the hardware implementation challenges of ZO optimization. Section IV presents the architectural design of the PGU within the proposed IPZO framework. Section V evaluates both the algorithmic effectiveness and hardware efficiency of the design, and Section VI concludes the work.

\section{Preliminaries}
\subsection{Spiking Transformers}
A standard transformer comprises multiple stacked blocks, each consisting of a multi-head self-attention (MHSA) module and a feedforward network (FFN), together with residual connections and layer normalization. In each MHSA module, the input representations are linearly projected into queries $Q$, keys $K$, and values $V$ using the weight matrices $W_Q, W_K,$ and $W_V$, respectively. Scaled dot-product attention is then computed as $A = \mathrm{softmax}(QK^{\top}/\sqrt{d_k})V$, where $d_k$ denotes the head dimension \cite{fournier2023practical}. The attention computation and the linear transformations within the FFN rely heavily on dense MVMs and typically require high-precision activations, posing considerable challenges for resource-constrained edge hardware.

Spiking transformers extend this architecture into the temporal domain by incorporating leaky integrate-and-fire (LIF) neurons into the MHSA and FFN modules \cite{cessac2011discrete, zhou2026spikingformer}. Following the linear projections, $Q$, $K$ and $V$ are encoded as temporal spike sequences $Q_t$, $K_t$, and $V_t$ over $T$ discrete time steps, where $t = 1, \ldots, T$. At each time step, attention is computed directly in the spike domain as $A_t=\operatorname{LIF}\!\left(\operatorname{LIF}\!\left(Q_tK_t^{\top}\right)V_t\right)$ \cite{10595893}. The intermediate attention scores $S_t=\operatorname{LIF}\!\left(Q_tK_t^{\top}\right)$ are regulated by the firing dynamics of LIF neurons instead of being normalized by softmax, thereby eliminating the hardware-intensive exponential and division operations required by conventional softmax attention.

\subsection{In-Memory Computing}
In-memory computing (IMC) alleviates the memory-wall bottleneck by performing computations directly within memory arrays. Depending on how the MAC operations are performed, IMC can be implemented in either the analog or the digital domain. Analog IMC typically activates multiple word lines (WLs) simultaneously and accumulates the resulting currents or charges along bit lines (BLs), producing MAC outputs in the analog domain. Digital IMC instead computes bitwise products using arithmetic logic embedded within the memory macro and accumulates the resulting partial products via digital circuits (e.g., adder trees), thereby generating digitally encoded MAC outputs \cite{chich202189}. Despite their different circuit implementations, both commonly employ a weight-stationary dataflow, in which weights remain stored within the array while input activations are streamed through it for computation.

IMC can be implemented using diverse memory technologies, each offering different trade-offs among density, reliability, and manufacturability. Among volatile memories, static random access memory (SRAM) is widely adopted for its compatibility with standard CMOS processes, mature foundry support, and robust operation, although its multi-transistor bit cell imposes considerable area overhead \cite{jhang2021challenges, kneip2021impact}. Emerging non-volatile memories (NVMs), including resistive random-access memory (ReRAM), phase change memory (PCM) and magnetic random-access memory (MRAM), enable higher storage density and compact crossbar integration. However, they remain subject to technology-dependent device nonidealities, including conductance drift, limited on/off ratios, device variability, and finite write endurance \cite{aguirre2024hardware, khaddam2021hermes, verma2023neuromorphic}. 

When mapped onto analog IMC implementations based on these memory technologies, spiking transformers benefit not only from inherent array-level parallelism but also from simplified computations enabled by binary, event-driven spike activations. In an IMC-based spiking transformer, input activations are represented as spike trains over time steps. Accordingly, the voltage applied to each WL switches between a fixed read voltage and ground, corresponding to spike values of `1' and `0', respectively. This binary gating reduces each computation from MAC to a pure accumulation, with only spike-activated rows contributing to the BL currents or charges. This sparsity-aware mechanism can improve energy efficiency while retaining the parallelism and compact hardware footprint offered by analog IMC.

\subsection{Zeroth-Order Optimization}
ZO optimization estimates gradients from finite differences of loss evaluations, making it applicable to non-differentiable objectives in which first-order derivatives are unavailable or prohibitively expensive to obtain \cite{chen2024deepzero}. By relying solely on forward-pass evaluations, ZO optimization eliminates the backward pass and avoids retaining intermediate activations for gradient computation \cite{suwandi2026breaking}. This advantage is particularly pronounced for SNNs, as BPTT must retain neuronal states at every time step, causing the activation memory to grow approximately linearly with the number of time steps $T$. Despite avoiding this activation-storage overhead, conventional ZO implementations still require storing a random perturbation vector whose dimensionality equals the total number of trainable model parameters. MeZO \cite{malladi2023fine} reduces this requirement by regenerating the perturbation vector from a shared random seed rather than explicitly storing it. The same vector can therefore be reproduced on demand during the positive and negative forward evaluations and the subsequent parameter update, bringing the memory consumption of ZO optimization close to that of inference.

For a model with parameters $\theta \in \mathbb{R}^d$ and a mini-batch $\mathcal{B} = \{x_1,\ldots,x_B\}$ of size $B$ sampled from the data distribution $\mathcal{D}$, let $\mathcal{L_{B}}(\theta)$ denote the empirical loss over $\mathcal{B}$. The gradient estimate $\hat{g}$ of ZO optimization is then given by
\begin{equation}
\hat{g} = \frac{1}{q} \sum_{i=1}^{q} \frac{\mathcal{L_{B}}(\theta + \epsilon z_i) - \mathcal{L_{B}}(\theta - \epsilon z_i)}{2\epsilon} z_i, 
\end{equation}
where $q$ denotes the number of random perturbations for each gradient estimation, $\epsilon > 0$ controls the perturbation magnitude, and $z_i \in \mathbb{R}^d$ is sampled from isotropic distributions such as the standard Gaussian $\mathcal{N}(0, \mathbf{I})$ or uniform distribution. The resulting gradient estimate is then used to update the model parameters according to the stochastic gradient descent (SGD) rule,
\begin{equation}
\theta_{t+1} = \theta_t - \eta \hat{g}, 
\end{equation}
where $\eta$ is the learning rate. Each optimization step therefore requires $2q$ forward-pass evaluations, corresponding to the positive and negative perturbations for each $z_i$.

\section{Challenges}
\subsection{Massive Random Number Generation}
Implementing ZO optimization requires an independently sampled perturbation element for every weight parameter at each optimization step, creating a substantial demand for on-chip random number generation. Although uniformly distributed perturbations have been shown to be an effective alternative to Gaussian-distributed perturbations \cite{tan2025perturbation, qinzeroth, liu2018zeroth}, their generation still incurs prohibitive hardware overhead in a fully parallel implementation. For a perturbation vector that matches the dimensionality of the weight matrix, the area required for the LFSR-based uniform RNGs comprising multiple flip-flops and feedback logic is much larger than that of the SRAM or NVM cells storing the corresponding weights. Consequently, this pronounced footprint mismatch between stochastic perturbation generation and static weight storage makes fully parallel per-weight RNGs impractical for large-scale IMC-based on-chip learning. 

To alleviate this footprint mismatch, perturbation-efficient ZO optimization (PeZO) \cite{tan2025perturbation} reuses the outputs of a small uniform RNG array across multiple weights, constructing full perturbation vectors through output concatenation and cyclically shifted remapping over time. Motivated by the low intrinsic dimensionality of language-model updates, PeZO effectively reduces the required number of RNGs. Nevertheless, the resulting statistical correlations among perturbation elements can degrade learning accuracy when insufficient random numbers are provided. Moreover, the RNG configuration required to avoid such degradation varies across models and tasks, complicating hardware provisioning and limiting portability across workloads. Consequently, reducing RNG overhead while maintaining sufficient perturbation diversity remains a critical challenge for scalable IMC-based ZO training.

\subsection{Frequent Read-Modify-Write Operations}
Although IMC architectures derive much of their efficiency from a weight-stationary dataflow, this advantage can be undermined in the on-chip implementation of ZO training. During inference, weights are written to the memory array once and remain stationary as successive input activations are processed, allowing the weight-loading cost to be amortized over all subsequent MVMs. During explicit-perturbation ZO training, however, each optimization step perturbs the model parameters, requiring the stored weight matrix to be repeatedly read from the array, modified, and written back. A cost incurred only once during inference therefore becomes a recurring per-step overhead during training. These repeated RMW operations not only consume considerable energy but also impose a throughput penalty because of sequential row-wise reads and writes. Furthermore, frequent reprogramming is particularly detrimental to NVM-based arrays as it exacerbates write-endurance limitations and shortens operational lifetime. As a result, explicitly injecting perturbations into the memory array disrupts the weight-stationary dataflow central to IMC efficiency, making repeated RMW operations a bottleneck for on-chip ZO training.

While PeZO reduces the hardware overhead of RNGs, it still explicitly applies perturbations to the stored weights, leaving frequent RMW operations largely unchanged. To mitigate the write-endurance constraints of NVMs, a dedicated eDRAM-IMC array has been introduced to store and accumulate the dynamic perturbations while retaining the weights in the RRAM crossbar \cite{chen2025analog}. Although this avoids repeatedly reprogramming the RRAM weights, the perturbations must still be read from and written to an additional memory array, shifting rather than eliminating the associated energy consumption. NoiseZO \cite{wang2025noisezo} instead maps the same weight matrix onto two RRAM crossbars and exploits the conductance differences induced by independent programming noise as perturbations. Although this approach avoids repeated RMW operations, it relies on device-specific write-noise characteristics, limiting its portability across memory technologies. Therefore, a solution that preserves weight stationarity without requiring dedicated perturbation storage across different IMC technologies is still lacking.

\section{Architecture and Design}
\subsection{Implicit-Perturbation ZO Architecture}
\begin{figure}[t]
  \centering
  \includegraphics[width=1\linewidth, trim={0 0.1cm 0 0}, clip]{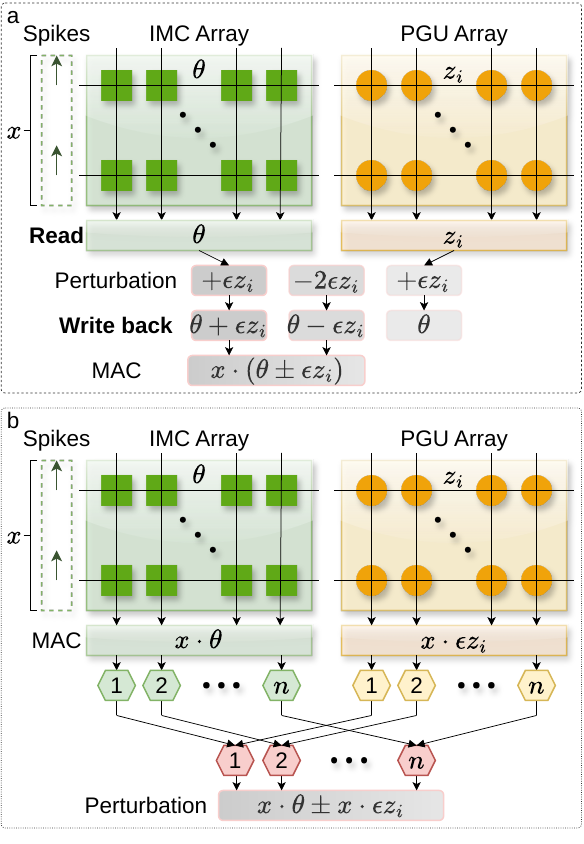}
  \caption{Comparison of (a) EPZO and (b) IPZO architectures for computing perturbed MVM outputs. The symbols $x$, $\theta$, $z_i$, and $\epsilon$ denote the input spike vector, weight matrix, $i$-th random perturbation vector, and perturbation scaling factor, respectively.}
  \label{fig:ipzo}
\end{figure}

In an IMC-based ZO accelerator, the random perturbation can be injected either onto the stored weights before MAC computation or onto the accumulated output afterward. Since ZO fundamentally perturbs the model weights, the most direct realization adopts the former, adding the perturbation to the stored weight matrix explicitly, as illustrated in Fig.~\ref{fig:ipzo}a. In this EPZO scheme, the weight matrix $\theta$ is first read from the IMC array while the PGU generates the random vectors $z_i$ in parallel. The weights are then perturbed by $+\epsilon z_i$ and written back as $\theta + \epsilon z_i$ for the positive evaluation. A subsequent $-2\epsilon z_i$ then flips them to $\theta - \epsilon z_i$ for the negative evaluation, and a final $+\epsilon z_i$ restores the array to $\theta$, a step that becomes indispensable under multiple perturbations. Driven by the input spike vector $x$, the MVM operation over the modified weights produces the perturbed output $x \cdot (\theta \pm \epsilon z_i)$ at the cost of three full-array RMW cycles per perturbation.

Rather than perturbing the stored weights explicitly, the proposed IPZO architecture injects the perturbation into the accumulated output, thereby preserving the weight stationarity of the IMC array. As illustrated in Fig.~\ref{fig:ipzo}b, the IMC array executes the standard MVM with the input spike vector $x$ to generate the unperturbed sum $x \cdot \theta$, while the PGU dynamically generates random perturbations and computes the scaled perturbation sum $x \cdot \epsilon z_i$ through a corresponding MVM. These two intermediate results are then merged to reconstruct the perturbed output, which is functionally equivalent to the EPZO baseline by the distributivity of the MVM,
\begin{equation}
x \cdot (\theta \pm \epsilon z_i) = x \cdot \theta \pm x \cdot \epsilon z_i.
\end{equation}
This separation allows IPZO to decouple the stochastic perturbation from the two-dimensional weight-matrix domain and remap it onto the one-dimensional accumulation-sum domain, leaving the stored weights untouched during a single optimization step.

\subsection{Event-Triggered Weight Perturbation}
\label{design}
\subsubsection{Dimension Reduction}
When deploying the IPZO architecture on SNNs, the PGU dimension can be reduced by leveraging the event-driven sparsity of spiking activations. At each time step, only the rows activated by input spikes contribute to the MVM, while the remaining rows remain inactive. The stochastic perturbation therefore only needs to be generated for the weights associated with these active rows rather than for the entire crossbar. This sparsity-induced decoupling allows the PGU row dimension to be determined by the number of concurrently active rows rather than the physical row dimension of the IMC array.

\begin{figure}[t]
  \centering
  \includegraphics[width=1\linewidth, trim={0 0.3cm 0.1cm 0.3cm}, clip]{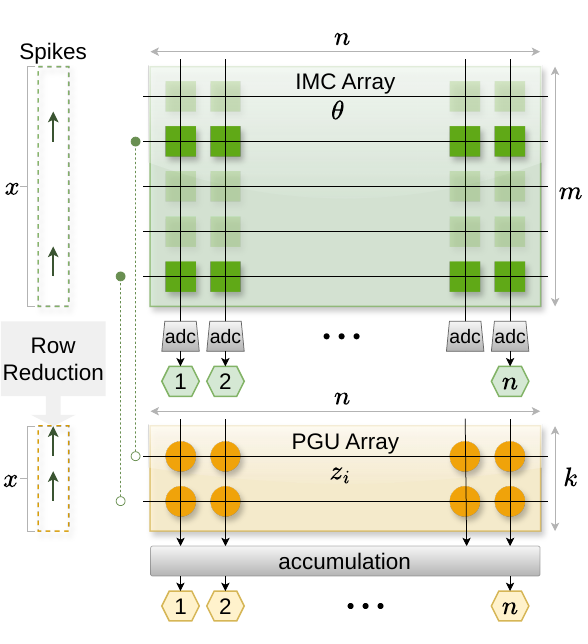}
  \caption{Block diagram of the sparse IPZO architecture with a row-dimension-reduced PGU array. $m$ and $n$ denote the numbers of rows and columns of the IMC array, and $k$ represents the number of PGU rows ($k \ll m$).}
  \label{fig:pgu1}
\end{figure}

Fig.~\ref{fig:pgu1} illustrates the resulting row-reduced PGU. Given an $m \times n$ IMC array storing $n$ multi-bit weights per row, the input spike vector $x$ activates $k$ rows on average at each time step, where $k \ll m$. The PGU is therefore provisioned with $k$ physical rows, which are dynamically mapped to the active IMC rows according to their spike addresses. When the number of active rows transiently exceeds this capacity because of sparsity fluctuations, the remaining active rows are processed over successive cycles. Therefore, this row-reduced design generates perturbations covering all weights participating in the MVM while substantially reducing the hardware overhead.

\subsubsection{Perturbation Independence}
\begin{figure}[t]
  \centering
  \includegraphics[width=1\linewidth, trim={0.1cm 0.2cm 0cm 0}, clip]{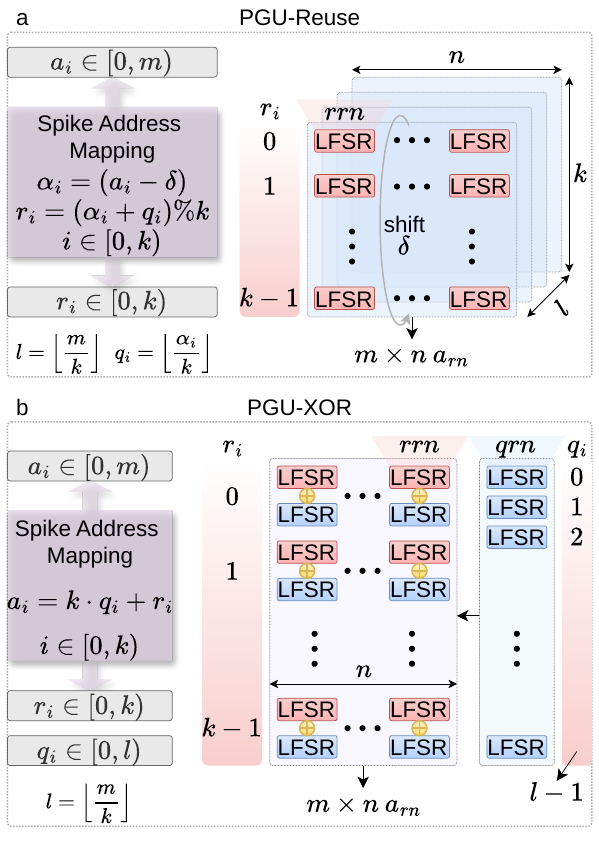}
  \caption{Perturbation generation schemes for the row-dimension-reduced PGU: (a) PGU-Reuse with group-based perturbation sharing across rows; (b) PGU-XOR with address decomposition and XOR-based combination of quotient- and remainder-associated LFSRs.}
  \label{fig:pgu2}
\end{figure}

While row-dimension reduction effectively minimizes the PGU hardware overhead, mapping the reduced-row PGU across the full weight matrix makes it challenging to ensure independence of perturbations. A baseline solution termed PGU-Reuse realizes this mapping by sharing the perturbation vectors generated by the reduced-row PGU across multiple weight rows. As shown in Fig.~\ref{fig:pgu2}a, PGU-Reuse partitions the $m$-row weight matrix into $l$ interleaved row groups ($l = \lfloor m/k \rfloor$), allowing multiple weight rows to share the same set of $k \times n$ perturbation vectors generated by the LFSR array. To randomize the perturbation sharing pattern across iterations, the spike address $a_i$ is first shifted by an iteration-dependent offset $\delta$, yielding an intermediate address $\alpha_i = a_i - \delta$. This intermediate address $\alpha_i $ is then mapped to a specific PGU row through the modulo operation $r_i = (\alpha_i + q_i) \pmod k$, where the quotient $q_i = \lfloor \alpha_i / k \rfloor$. Although this dynamic remapping changes the assignment of shared perturbation vectors across weight rows over different iterations, the repeated reuse of the same random number bank inevitably introduces statistical correlations.

To address this, we introduce a new PGU-XOR scheme through address-driven XOR recombination, enabling statistically independent perturbations across the entire weight matrix. As shown in Fig.~\ref{fig:pgu2}b, the address-driven XOR recombination first decomposes each spike address $a_i \in [0, m)$ into a quotient $q_i \in [0, l)$ and a remainder $r_i \in [0, k)$ such that $a_i = k \cdot q_i + r_i$. When $k$ is a power of two, $q_i$ and $r_i$ are extracted directly from the high- and low-order address bits, eliminating the need for a hardware divider. Therefore, the PGU-XOR comprises two independent LFSR groups including $l$ quotient-associated LFSRs (Q-LFSRs) and a $k \times n$ array of remainder-associated LFSRs (R-LFSRs), which generate the corresponding random numbers $qrn$ and $rrn$, respectively. For a given spike address $a_i$, the single $qrn$ selected by $q_i$ is bitwise XORed with the corresponding row of $rrn$ indexed by $r_i$, producing the $n$-element perturbation vector associated with that address. Because every Q-LFSR and R-LFSR is configured with a distinct primitive polynomial, the resulting quotient–remainder XOR combinations generate statistically independent perturbation vectors for all $m$ spike addresses. Consequently, PGU-XOR eliminates the perturbation correlations introduced by PGU-Reuse, thereby restoring the perturbation diversity required for effective ZO gradient estimation. 

\subsubsection{Accumulation Contention}
\begin{figure}[t]
  \centering
  \includegraphics[width=1\linewidth, trim={0 0.3cm 0 0}, clip]{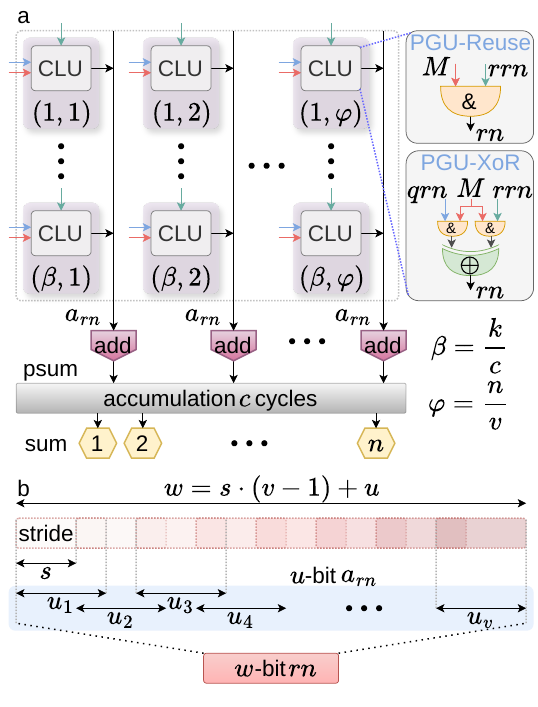}
  \caption{(a) Circuit diagram of the perturbation accumulation architecture with a $\beta \times \varphi$ CLU array for PGU-Reuse and PGU-XOR, requiring $c=k/\beta$ accumulation cycles. (b) Overlapping bit-slicing scheme using stride $s$ to partition the $w$-bit CLU output $rn$ into $v$ pseudo-random segments.}
  \label{fig:pgu3}
\end{figure}

While XOR recombination restores perturbation independence, it forces multiple spike addresses to share access to the Q-LFSR and R-LFSR resources. Under multi-row concurrent activation, addresses that decompose to the same remainder contend for the same R-LFSR row within a single cycle, creating row-level resource contention along the XOR-and-accumulation datapath. Rather than over-provisioning the LFSR resources, we resolve this contention with an accumulation network (Fig.~\ref{fig:pgu3}a) that serializes the contending accesses across $c$ cycles, trading a bounded latency for a compact hardware footprint.

This accumulation network is organized as a $\beta \times \varphi$ array of combinational logic units (CLUs), where $\beta = k/c$ and $\varphi = n/v$ are set by the accumulation cycle count $c$ and the segmentation number $v$. Within each CLU of PGU-XOR, the $qrn$ and $rrn$ selected by the corresponding $q_i$ and $r_i$ are first gated by a mask $M$ through AND operations, passing only those associated with valid spikes so that the network adapts to a varying number of active rows ($ \le \beta $) per cycle. The gated $qrn$ and $rrn$ are then combined by an XOR to produce the wide $w$-bit random number $rn$. PGU-Reuse follows the same datapath but bypasses the XOR, directly forwarding the gated $rrn$ as $rn$. As illustrated in Fig.~\ref{fig:pgu3}b, each $w$-bit $rn$ is partitioned into $v$ overlapping $u$-bit segments $a_{rn}$ with a stride of $s$ bits, where $w=s \cdot (v-1)+u$. This overlapping bit-slicing scheme improves the utilization of every generated bit. Finally, these segments $a_{rn}$ aligned to the same column are summed by the downstream adders and accumulated over $c$ cycles to yield the perturbation sum.

The accumulation cycle count $c$ serves as a configurable design parameter that balances hardware area against accumulation latency. A larger $c$ shrinks the CLU array ($\beta=k/c$) but distributes the random-number generation and accumulation over more cycles. This multi-cycle accumulation, however, typically does not introduce a system-level latency penalty within the IPZO architecture. Since the IMC pipeline inherently spends multiple clock cycles on MAC operations, column-level analog-to-digital conversion and partial-sum aggregation, the PGU accumulation is overlapped within these cycles as long as $c$ stays within the IMC pipeline latency. This relaxed timing allows $c$ to be configured according to the available hardware budget while keeping the accumulation latency effectively hidden.

\subsection{Data Flow}
\begin{figure*}[htbp]
  \centering
  \includegraphics[width=0.80\textwidth, trim={0 0.3cm 0 0.1cm}, clip]{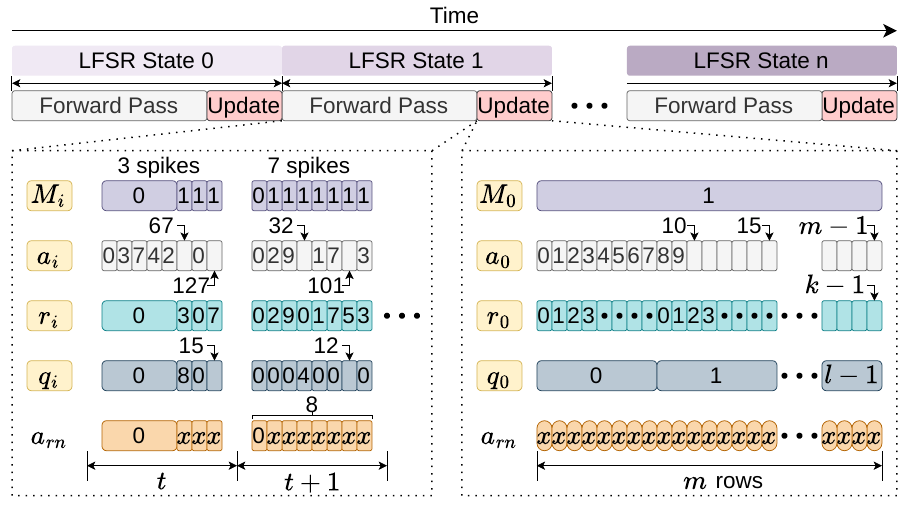}
  \caption{Data flow of the PGU-XOR across forward pass and parameter update phases.}
  \label{fig:flow}
\end{figure*}

Fig.~\ref{fig:flow} illustrates the data flow of the PGU-XOR across the alternating forward-pass and weight-update phases. Following the ZO training framework, PGU-XOR generates perturbations for gradient estimation during the forward pass and reproduces the identical perturbations for the weight parameter update afterward. This reproducibility is achieved by holding the LFSR states constant throughout a single optimization iteration, so that the same random numbers can be regenerated on demand without being stored.

The two phases, however, differ in their perturbation generation process. During the forward pass, only the spike-activated rows participate in the computation and perturbations are therefore generated exclusively for these active rows. As illustrated in Fig.~\ref{fig:flow}, three spikes are active at time step $t$, causing the mask $M_i$ to assert its three least significant bits. The resulting addresses $a_i$ are then decomposed into ($r_i$, $q_i$) pairs to generate the associated perturbation segments $a_{rn}$. In contrast to the event-driven forward pass, the update phase sequentially traverses the entire weight matrix to regenerate perturbations for every row. Accordingly, a single bit of the mask $M_0$ is asserted while the address $a_0$ increments sequentially from 0 to $m-1$. As $a_0$ advances, $r_0$ cycles repeatedly from 0 to $k-1$ whereas $q_0$ increments every $k$ cycles. This sequential traversal ensures that the perturbations used during parameter updates remain identical to those employed for gradient estimation.

\begin{figure}[t]
  \centering
  \includegraphics[width=1\linewidth, trim={0.25cm 0.2cm 0.2cm 0.1cm}, clip]{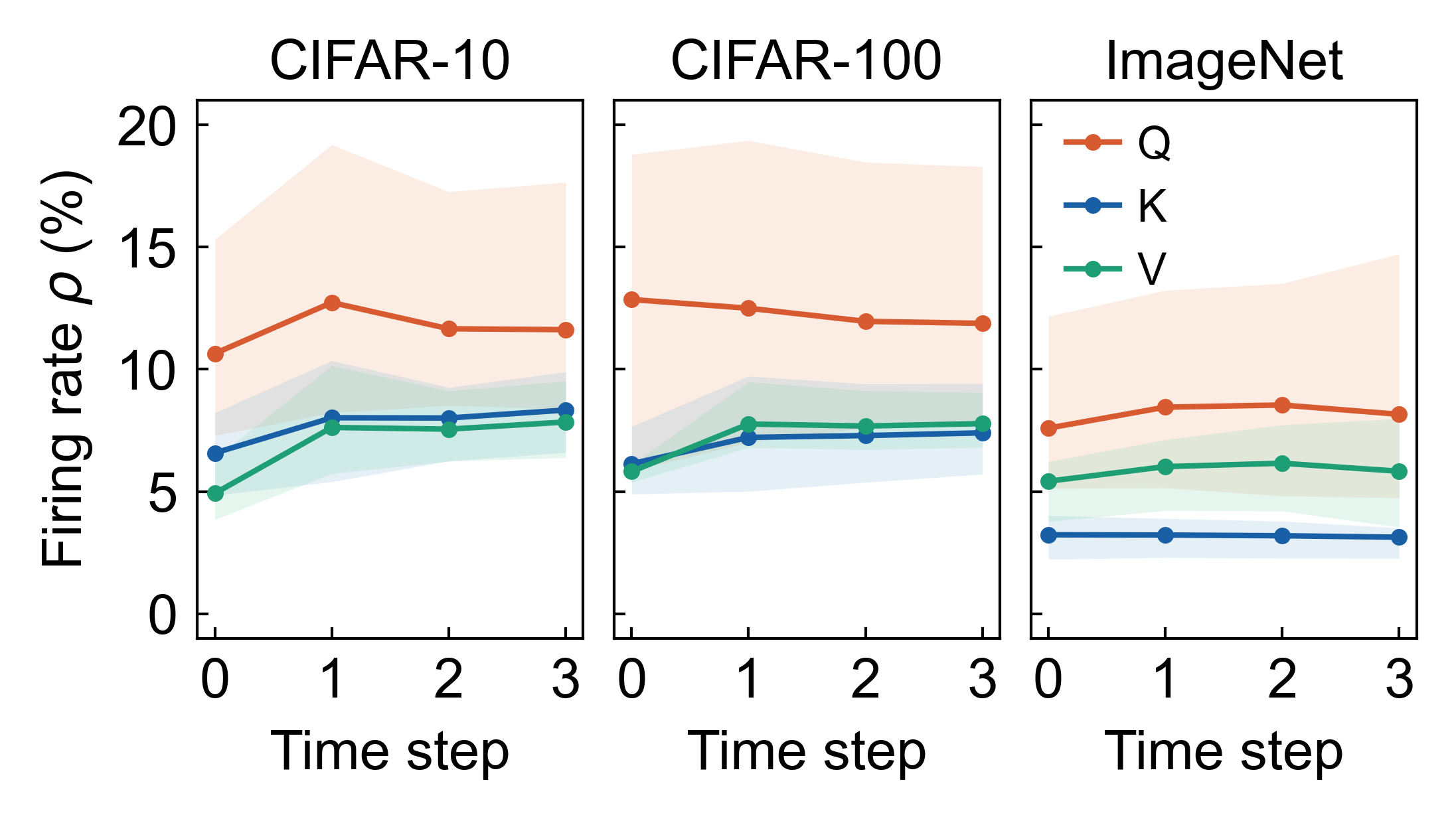}
  \caption{Mean firing rate $\rho$ of Q, K, and V LIF neurons across transformer blocks at each time step. Shaded regions indicate its range across transformer blocks.}
  \label{fig:sparsity}
\end{figure}

\section{Experiments and Results}
\subsection{Spatial Sparsity Analysis}
Fig.~\ref{fig:sparsity} presents the mean firing rates $\rho$ of Q, K, and V LIF neurons in Spikingformer \cite{zhou2026spikingformer} at each time step for three representative datasets. The shaded regions indicate the variation across multiple transformer blocks. Across all three datasets, the Q, K and V LIF neurons exhibit consistently sparse firing activity throughout the inference. The mean firing rates $\rho$ of all three neuron types, averaged across transformer blocks, remain below $15$\% at every time step. Among them, the Q LIF neurons exhibit the highest firing activity, with block-wise maxima reaching approximately $20$\% for CIFAR-10 and CIFAR-100 and $15$\% for ImageNet, while the K and V LIF neurons remain predominantly below $10$\%. These consistently low firing rates demonstrate the inherent sparsity of spiking activity and justify the row-dimension reduction adopted in the PGU design. Guided by this observation, both PGU-XOR and PGU-Reuse are configured with $k = 8$ physical rows for an $m \times n = 128 \times 16$ IMC array, providing a single-cycle capacity of $6.25$\%. Each of the 16 columns stores an 8-bit weight, resulting in 128 bit-columns at the cell level. Rather than provisioning additional physical rows for firing rates above $6.25$\%, multi-cycle scheduling is employed to trade latency for area. For an instantaneous firing rate $\rho$, the required number of scheduling cycles is $N_c=\lceil m\rho/k \rceil $. With $m=128$ and $k=8$, four scheduling cycles can accommodate firing rates of up to $25$\%, exceeding the maximum of approximately $20$\% observed in Fig.~\ref{fig:sparsity}. These additional cycles can be largely overlapped within the IMC pipeline, thereby limiting their impact on system-level latency.

\subsection{Statistical Property Evaluation}
Using the $k=8$ configuration for a $128 \times 16$ weight matrix, we evaluate the statistical properties of perturbations generated by PGU-Reuse and PGU-XOR in terms of bit-level uniformity, temporal randomness, and spatial independence. In this setup, PGU-Reuse employs an $8 \times 2$ array of R-LFSRs, while PGU-XOR additionally incorporates 16 Q-LFSRs, each implemented as a 36-bit ($w=36$) LFSR with a distinct primitive polynomial. Through overlapping bit slicing and quotient–remainder address expansion, these reduced LFSR resources are logically expanded to generate perturbations covering the entire $128 \times 16$ weight matrix. 

\begin{figure}[t]
  \centering
  \includegraphics[width=1\linewidth, trim={0.25cm 0.25cm 0.2cm 0.1cm}, clip]{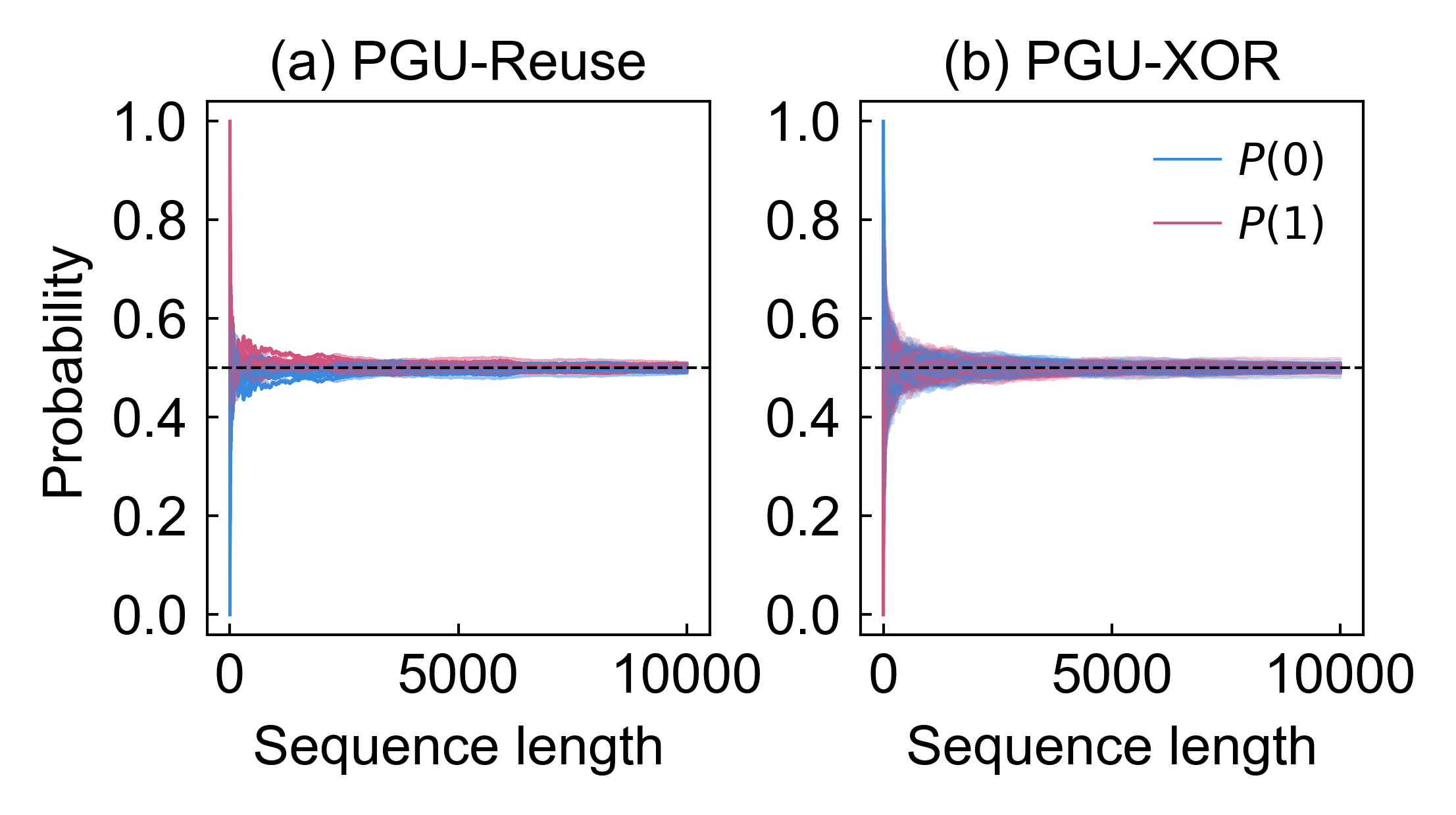}
  \caption{Evolution of bit probability distribution in 128 random sequences $rn$ generated by (a) PGU-Reuse and (b) PGU-XOR.}
  \label{fig:bit_prob}
\end{figure}

We first examine bit-level uniformity, which requires a balanced distribution of bits 0 and 1 in each generated sequence. Fig.~\ref{fig:bit_prob} presents the evolution of the bit probabilities for the 128 perturbation sequences ($rn$) associated with the first column of the weight matrix, each extended to a length of 10,000. For both configurations, the probabilities of bit 0 ($P(0)$) and bit 1 ($P(1)$) rapidly converge toward 0.5 as the sequence length grows, confirming that both PGU-Reuse and PGU-XOR produce uniform binary sequences. Bit-level uniformity therefore does not distinguish the two architectures, motivating a further evaluation of their temporal and spatial statistical properties.

\begin{figure}[t]
  \centering
  \includegraphics[width=1\linewidth, trim={0 0.5cm 0 0}, clip]{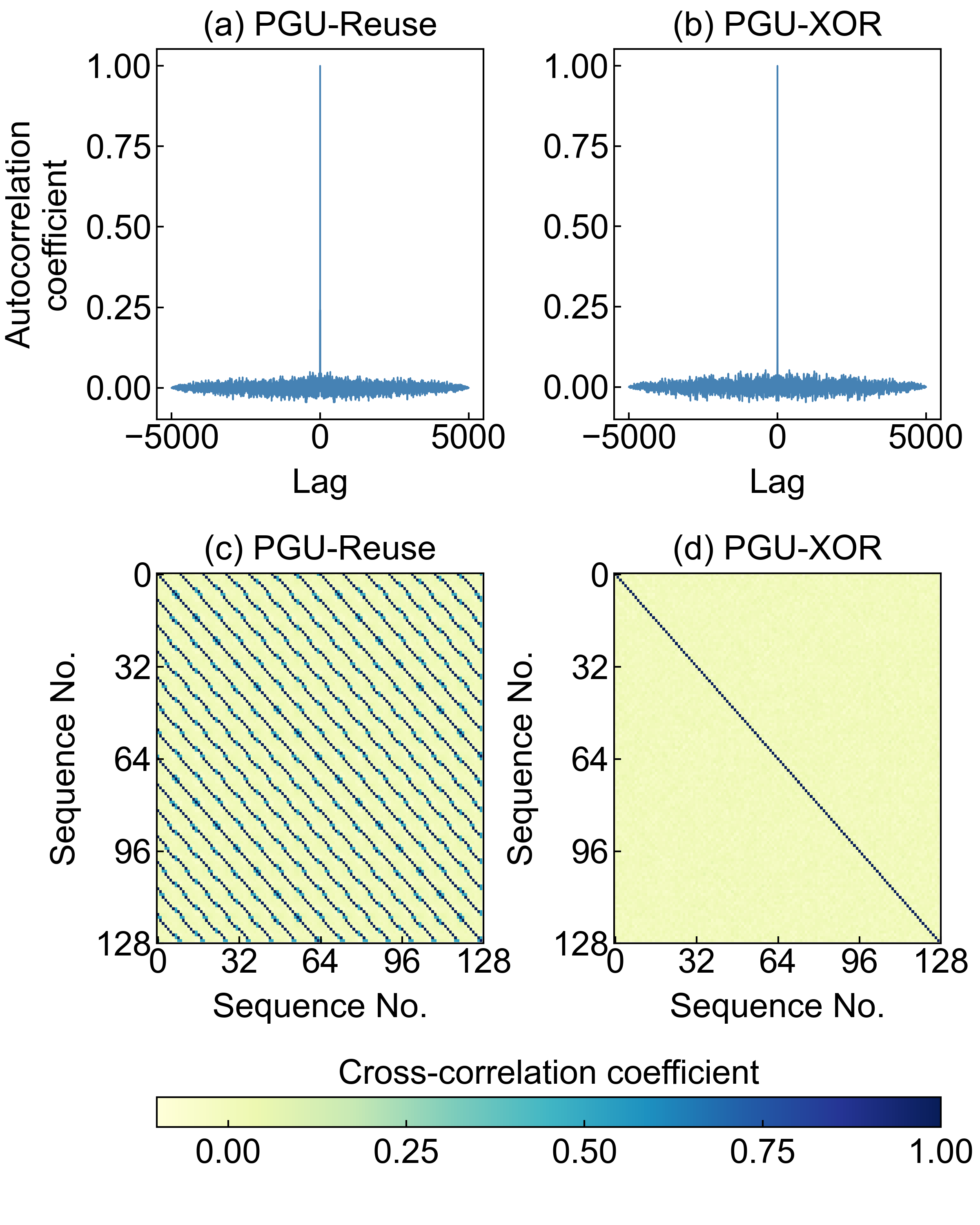}
  \caption{Statistical correlation properties of random sequences $rn$: autocorrelation coefficient of a single sequence generated by (a) PGU-Reuse and (b) PGU-XOR; cross-correlation matrix among 128 sequences generated by (c) PGU-Reuse and (d) PGU-XOR.}
  \label{fig:auto_cross_corr}
\end{figure}

Figs.~\ref{fig:auto_cross_corr}a and b show the autocorrelation of a single sequence generated by PGU-Reuse and PGU-XOR, respectively. In both cases, the autocorrelation peaks only at zero lag due to the self alignment, whereas the coefficients across all non-zero lags remain close to zero, indicating negligible temporal self-correlation. The distinction emerges in the spatial domain, where the cross-correlation among the 128 sequences differs sharply between the two architectures. As shown in Fig.~\ref{fig:auto_cross_corr}c, PGU-Reuse exhibits pronounced periodic cross-correlations among different perturbation sequences. These structured correlations arise from the repeated reuse of shared LFSR states across parallel rows under the fixed cyclic remapping scheme. In contrast, the cross-correlation matrix of PGU-XOR (Fig.~\ref{fig:auto_cross_corr}d) remains close to zero except along the self-correlation diagonal, indicating that the 128 sequences are mutually uncorrelated. By eliminating structured spatial correlations, PGU-XOR restores statistically independent perturbations across all weight rows, thereby preserving the isotropic perturbation directions required for unbiased ZO gradient estimation.

\begin{figure}[t]
  \centering
  \includegraphics[width=1\linewidth, trim={0 0.5cm 0 0}, clip]{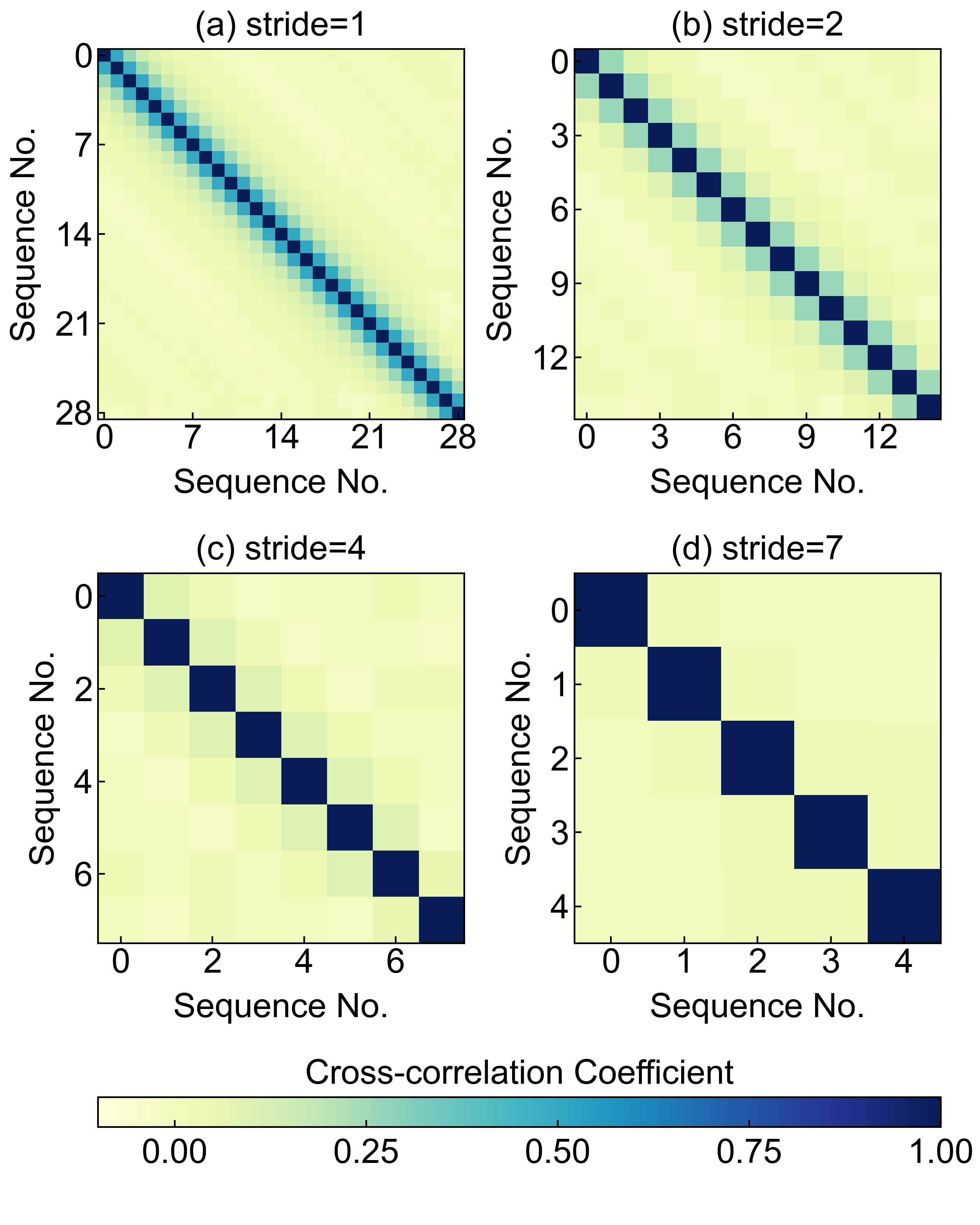}
  \caption{Cross-correlation characteristics of 8-bit ($u$) random sequences $a_{rn}$ extracted from a 36-bit ($w$) PGU-XOR sequence with different stride configurations: (a) $s=1$ with 29 sequences; (b) $s=2$ with 15 sequences; (c) $s=4$ with 8 sequences; (d) $s=7$ with 5 sequences.}
  \label{fig:corr_stride}
\end{figure}

The overlapping bit-slicing scheme introduces an additional source of correlation between the generated perturbation segments. Because adjacent segments share overlapping bits, a smaller stride inevitably leads to stronger inter-segment correlation. We therefore evaluate the influence of the stride $s$ on the cross-correlation among the partitioned 8-bit ($u=8$) perturbation segments $a_{rn}$, generated from a 36-bit ($w=36$) LFSR output. For narrow strides of $s=1$ ($v=29$, Fig.~\ref{fig:corr_stride}a) and $s=2$ ($v=15$, Fig.~\ref{fig:corr_stride}b), the resulting segments exhibit pronounced cross-correlations exceeding 0.5 between adjacent segments along the main diagonal. Increasing the stride to $s=4$ ($v=8$, Fig.~\ref{fig:corr_stride}c) reduces the maximum cross-correlation between distinct segments below 0.1, while a further increase to $s=7$ ($v=5$, Fig.~\ref{fig:corr_stride}d) suppresses the off-diagonal correlations to nearly zero. These results reveal a clear trade-off between statistical independence and hardware efficiency. Larger strides effectively reduce inter-segment correlation but simultaneously decrease the number of available perturbation segments. We therefore adopt $s=4$, which maintains low inter-segment correlation while preserving a sufficient segment count ($v=8$).

\subsection{Accuracy Evaluation}
\label{accuracy}
\begin{figure}[t]
  \centering
  \includegraphics[width=1\linewidth, trim={0.25cm 0.3cm 0.2cm 0.2cm}, clip]{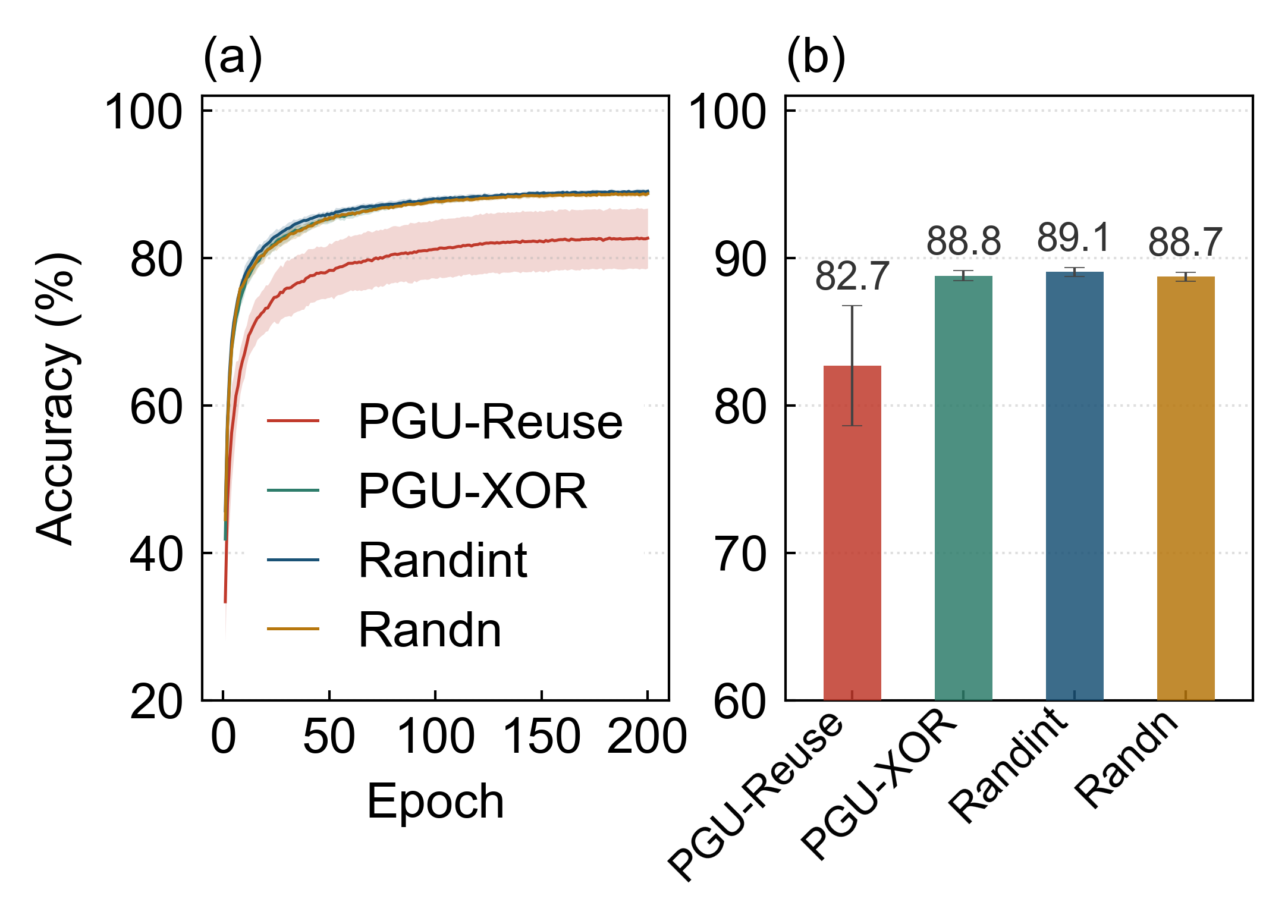}
  \caption{Accuracy of a three-layer SNN-MLP (1024-128-10) trained from scratch on MNIST with MeZO under four perturbation schemes including PGU-Reuse, PGU-XOR, Randint, and Randn: (a) accuracy versus training epoch and (b) final accuracy and standard deviation over 10 independent runs.}
  \label{fig:accuracy}
\end{figure}

The training performance of a ZO-based SNN is evaluated using different perturbation generators, including the hardware implementations PGU-Reuse and PGU-XOR, as well as two software references, namely uniformly distributed integer perturbations (\texttt{Randint}) and normally distributed perturbations (\texttt{Randn}). To ensure a fair comparison, the 8-bit integer perturbations ($\in [-128, 128)$) generated by PGU-Reuse, PGU-XOR and \texttt{Randint} are normalized to $[-0.5, 0.5)$ and scaled by $\sqrt{12}$ to match the unit variance of \texttt{Randn}. Experiments are conducted on a three-layer SNN-based multilayer perceptron (MLP) (1024-128–10) trained on MNIST, with ten independent runs from different random initializations for each perturbation scheme. As shown in Fig.~\ref{fig:accuracy}a, PGU-XOR exhibits nearly identical convergence behavior to both software references (\texttt{Randint} and \texttt{Randn}), with all three achieving stable optimization and narrow variance envelopes across ten independent runs. In contrast, PGU-Reuse exhibits substantially larger run-to-run variability, as reflected by its considerably wider error band. This is further confirmed by the final test accuracies in Fig.~\ref{fig:accuracy}b. PGU-XOR, \texttt{Randint}, and \texttt{Randn} reach comparable mean accuracies of 88.8\%, 89.1\%, and 88.7\% with tightly bounded standard deviations of 0.34\%, 0.31\%, and 0.30\%, respectively, whereas PGU-Reuse attains only 82.7\% with a substantially elevated deviation of 4.10\%. This degradation directly reflects the structured spatial correlations identified in Fig.~\ref{fig:auto_cross_corr}c. By reusing random perturbations across the rows of each weight matrix, PGU-Reuse violates the isotropy assumption underlying ZO gradient estimation, so the perturbations span far fewer independent directions than the dimension of the parameter space. 

\begin{figure}[t]
  \centering
  \includegraphics[width=1\linewidth, trim={0.24cm 0.3cm 0.2cm 0.2cm}, clip]{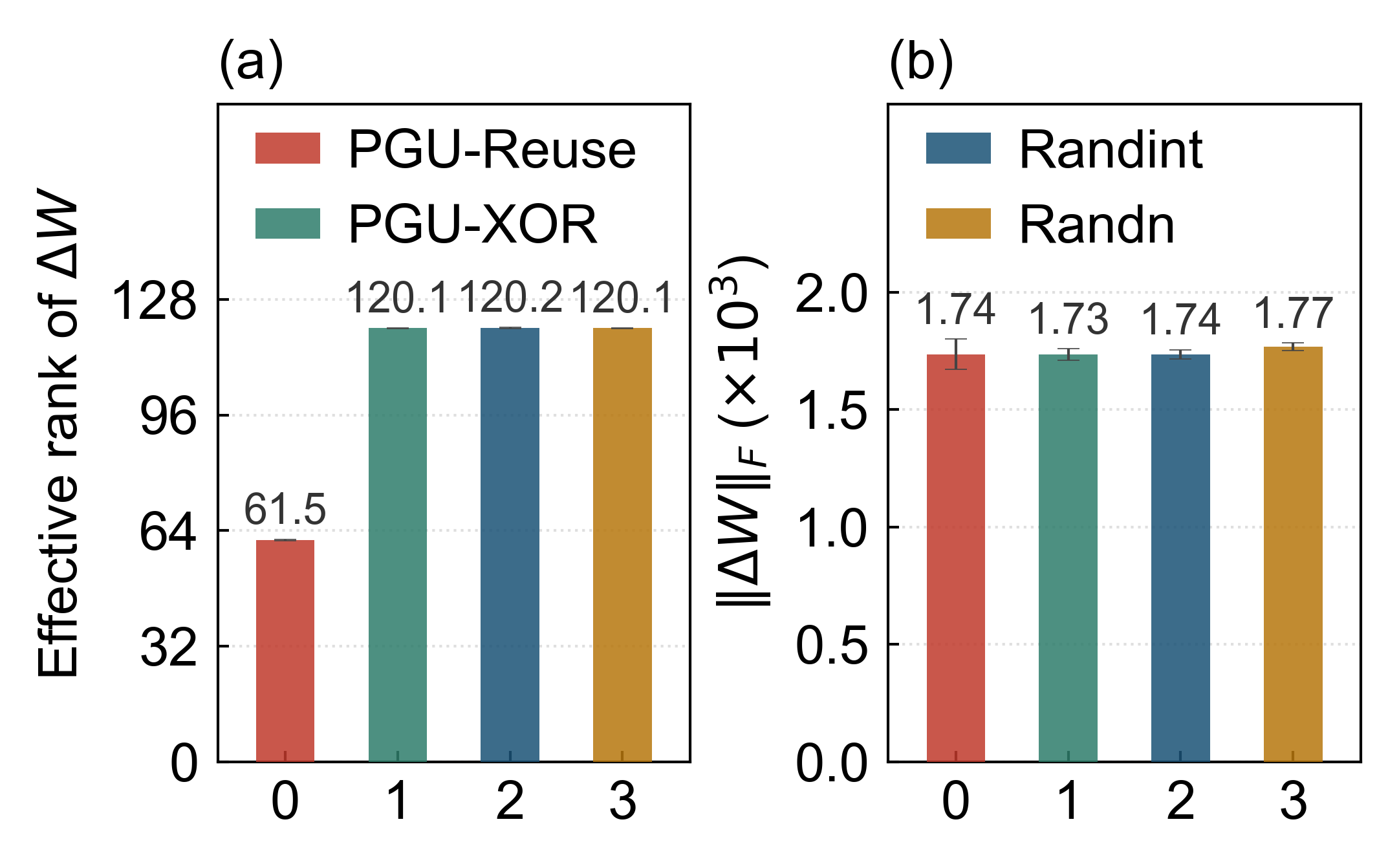}
  \caption{(a) Effective rank and (b) Frobenius norm $\|\Delta W\|_F$ of the weight change $\Delta W = W_{\mathrm{final}} - W_{\mathrm{init}}$ for the first layer of SNN-MLP trained from scratch using MeZO under four perturbation schemes including PGU-Reuse, PGU-XOR, Randint, and Randn.}
  \label{fig:rank}
\end{figure}

This loss of directional diversity is quantified by the effective rank of the first-layer weight update $\Delta W = W_{\mathrm{final}} - W_{\mathrm{init}}$ accumulated over training, which is bounded above by $m=128$. As shown in Fig.~\ref{fig:rank}a, PGU-XOR, \texttt{Randint}, and \texttt{Randn} all reach a near-full rank of about 120, indicating that the optimization explores a broad set of independent directions. PGU-Reuse, in contrast, collapses to 61.5, roughly half that of the other schemes. The 128 perturbation rows of PGU-Reuse are produced by cyclically shifting 8 rows of base random sequences into 16 groups, and because the shifts repeat every 8 groups, the last 8 groups merely duplicate the first 8. The row space of the perturbation matrix is therefore effectively capped at $m/2=64$, closely matching the measured 61.5 and confining every update to the same low-dimensional subspace. As shown in Fig.~\ref{fig:rank}b, the Frobenius norm $\|\Delta W\|_F$ is nearly identical across all four schemes at about $1.74 \times 10^{3}$. PGU-Reuse perturbs the weights by the same overall magnitude as the others, but concentrates this displacement in half as many independent directions. This restriction also accounts for the elevated run-to-run variability (Fig.~\ref{fig:accuracy}a), since the alignment between the gradient directions and the low-dimensional subspace varies with the random initialization. By decorrelating the perturbation rows, PGU-XOR removes this duplication and restores near-full-rank exploration of the parameter space, which underlies its accuracy and stability on par with the software references.

\begin{table}[t]
\centering
\caption{Fine-tuning Performance Using MeZO Under Different Random Numbers.}
\label{tab:mezo_results}
\setlength{\tabcolsep}{4.5pt}
\begin{tabular}{l cc cc cc}
\toprule
 & \multicolumn{2}{c}{Spikingformer \cite{zhou2026spikingformer}}
 & \multicolumn{4}{c}{SpikeGPT \cite{zhu2024spikegpt}} \\
\cmidrule(lr){2-3} \cmidrule(lr){4-7}
 & \multicolumn{2}{c}{CIFAR-10}
 & \multicolumn{2}{c}{WikiText-2}
 & \multicolumn{2}{c}{WikiText-103} \\
\cmidrule(lr){2-3} \cmidrule(lr){4-5} \cmidrule(lr){6-7}
 & Acc.\,(\%) & Epochs\textsuperscript{\dag}
 & PPL & Steps\textsuperscript{\dag}
 & PPL & Steps\textsuperscript{\dag} \\
\midrule
PGU-Reuse           & 66.85 & 247 & 66.01 & 3944 & 94.73 & 4000 \\
PGU-XOR             & 76.41 & \phantom{0}84 & 54.20 & \phantom{0}687 & 83.98 & \phantom{0}777 \\
\midrule
\texttt{Randint}             & 76.53 & \phantom{0}80 & 53.23 & \phantom{0}633 & 82.67 & \phantom{0}641 \\
\texttt{Randn}               & 76.56 & \phantom{0}83 & 53.07 & \phantom{0}653 & 82.32 & \phantom{0}742 \\
\bottomrule
\end{tabular}
\\[3pt]
\parbox{\linewidth}{\footnotesize \textsuperscript{\dag}Training epochs (Spikingformer) or steps (SpikeGPT) required to reach the final performance of PGU-Reuse.}
\end{table}

The impact of this rank deficiency is further examined on two large-scale spiking models including Spikingformer \cite{zhou2026spikingformer} and SpikeGPT \cite{zhu2024spikegpt}, fine-tuned with ZO optimization under the same four perturbation schemes. A 66.34M-parameter Spikingformer pre-trained on ImageNet-1K is fine-tuned on CIFAR-10 with 75\% of its parameters frozen, leaving 16.6M trainable parameters, whereas a 216M-parameter SpikeGPT pre-trained on OpenWebText2 is fully fine-tuned and evaluated on WikiText-2 and WikiText-103. 

The results are summarized in Table~\ref{tab:mezo_results}. Across all three tasks, PGU-XOR consistently matches the optimization performance achieved with the software perturbation generators (\texttt{Randint} and \texttt{Randn}), whereas PGU-Reuse exhibits a clear performance degradation. On CIFAR-10, PGU-XOR achieves $76.41$\% accuracy, closely matching \texttt{Randint} ($76.53$\%) and \texttt{Randn} ($76.56$\%), whereas PGU-Reuse reaches only $66.85$\%, $9.56$ percentage points lower than PGU-XOR. A similar trend holds on WikiText-2 and WikiText-103, where PGU-XOR attains perplexities (PPLs) of $54.20$ and $83.98$, respectively, close to the software references, while PGU-Reuse yields higher values of $66.01$ and $94.73$. With the final performance of PGU-Reuse as a common reference, PGU-XOR reaches the same accuracy on CIFAR-10 in $84$ epochs compared with $247$ epochs for PGU-Reuse, reducing training epochs by a factor of about $2.9$. On WikiText-2 and WikiText-103, PGU-XOR reaches the corresponding reference PPLs in $687$ and $777$ steps against $3944$ and $4000$ steps, corresponding to $5.7\times$ and $5.1\times$ fewer iterations, respectively. These results demonstrate that the spatial correlations introduced by PGU-Reuse impair convergence speed and degrade the final accuracy and perplexity at large model scales, whereas PGU-XOR preserves performance comparable to software perturbation generators across both vision and language tasks.

\subsection{Hardware Implementation}
PGU-Reuse and PGU-XOR are implemented in TSMC 16-nm CMOS technology using a standard-threshold-voltage digital cell library. Both designs support configurable 1-bit (BIT1) Rademacher ($\pm 1$) and 8-bit (BIT8) signed perturbation generation for a $128 \times 16$ IMC array. All hardware results reported in this section are obtained from post-layout implementations, with throughput evaluated at a target frequency $f$ of 1~GHz and power evaluated at the typical-typical (TT) process corner, $0.8$ V supply voltage, and $25~^\circ\mathrm{C}$. The following subsections evaluate their area, throughput, and energy consumption.

\subsubsection{Area and Throughput}
\begin{figure}[t]
  \centering
  \includegraphics[width=1\linewidth, trim={0.1cm 0.1cm 0.1cm 0.1cm}, clip]{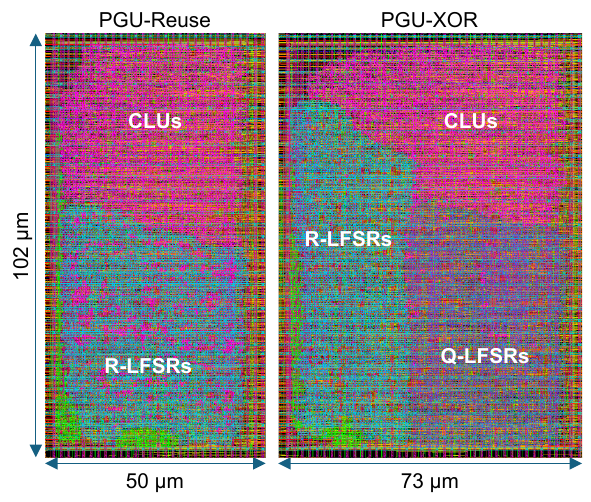}
  \caption{Layout comparison of the PGU-Reuse and PGU-XOR implementations in 16-nm CMOS technology under accumulation cycles of $c=8$.}
  \label{fig:layout}
\end{figure}

\begin{figure}[t]
  \centering
  \includegraphics[width=1\linewidth, trim={0.2cm 0.2cm 0.2cm 0.2cm}, clip]{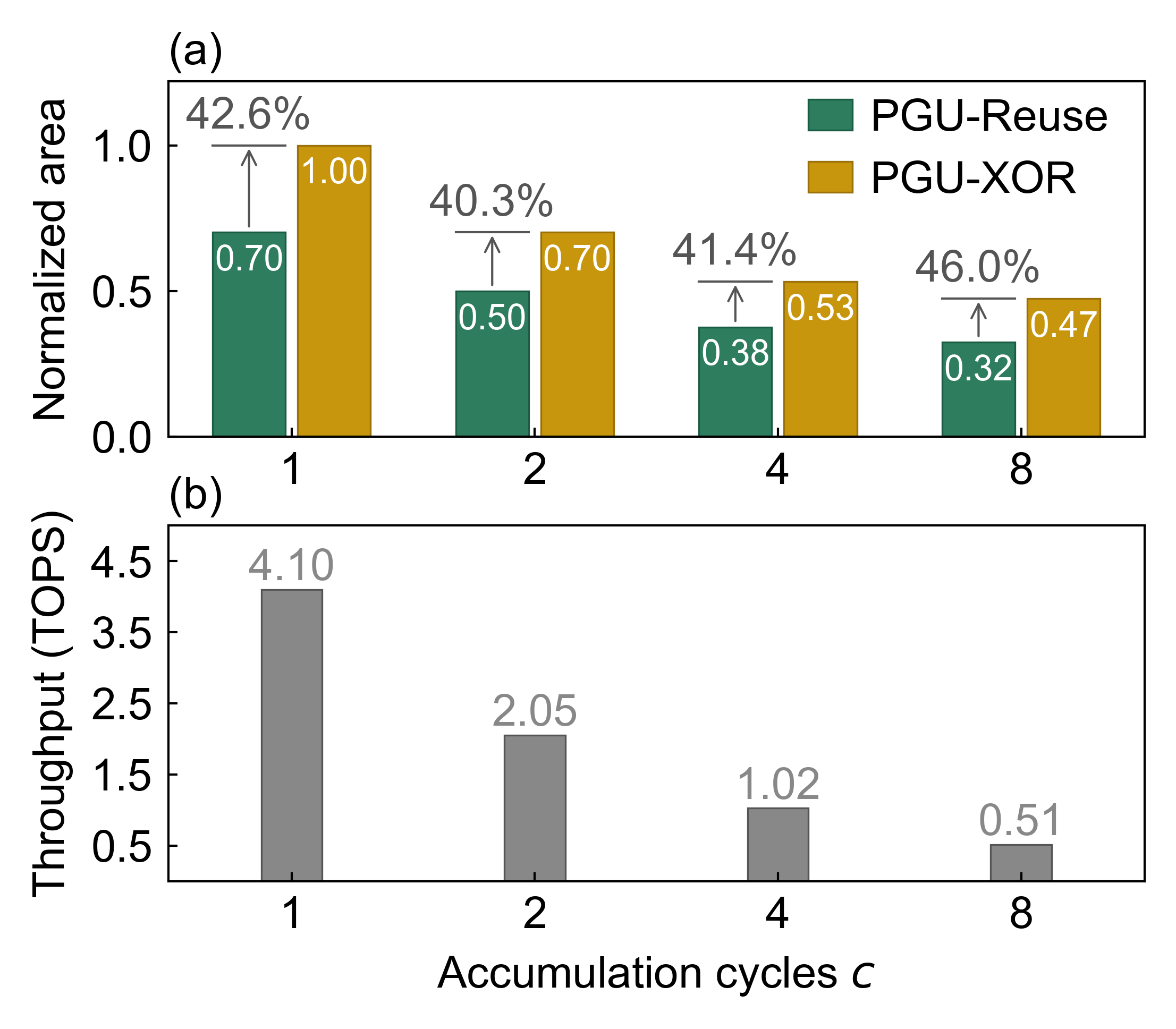}
  \caption{(a) Normalized post-layout area and (b) throughput of PGU-Reuse and PGU-XOR under different accumulation cycles $c$ in 16-nm CMOS technology. All designs are placed at $\sim 66\%$ cell utilization. One operation (OP) denotes one multiplication or one addition.}
  \label{fig:area}
\end{figure}

Fig.~\ref{fig:layout} compares the post-layout implementations of PGU-Reuse and PGU-XOR under the most area-efficient configuration ($c=8$), corresponding to a single-row CLU array ($\beta = 1$). The layouts are partitioned into color-coded functional regions, including the CLUs, the R-LFSRs, and the Q-LFSRs exclusive to PGU-XOR. Under a cell utilization of approximately 66\%, PGU-Reuse occupies $50\ \mu\text{m} \times 102\ \mu\text{m}$ ($\approx 5{,}100\ \mu\text{m}^2$), whereas PGU-XOR occupies $73\ \mu\text{m} \times 102\ \mu\text{m}$ ($\approx 7{,}446\ \mu\text{m}^2$), corresponding to a 46.0\% area overhead introduced by the additional Q-LFSRs for XOR-based perturbation generation. Beyond this single configuration, Fig.~\ref{fig:area}a evaluates the post-layout area for $c \in \{2, 4, 8\}$ with all values normalized to the largest configuration ($c=1$, PGU-XOR). Increasing $c$ progressively reduces the size of the CLU array through $\beta = k/c$, lowering the normalized area from 1.00 to 0.47 for PGU-XOR and from 0.70 to 0.32 for PGU-Reuse, while the overhead of PGU-XOR over PGU-Reuse stays within 40.3\%--46.0\% across all configurations. 

Despite the additional hardware required for perturbation decorrelation, the area of PGU-XOR at $c=1$ remains comparable to that of a representative $128 \times 128$ IMC macro, which can represent an 8-bit realization of the same $128 \times 16$ weight matrix targeted by the PGU-XOR. Such a referenced IMC macro occupies $0.124\ \text{mm}^2$ in 65-nm CMOS technology \cite{sharma2022reconfigurable}, corresponding to $\sim 16{,}000\ \mu\text{m}^2$ after scaling to 16-nm technology \cite{stillmaker2017scaling}. As $c$ increases, the relative area cost of PGU-XOR with respect to the IMC macro decreases further.

Fig.~\ref{fig:area}b presents the perturbation throughput of both PGU-Reuse and PGU-XOR under different accumulation cycles. Since PGU-XOR introduces the additional Q-LFSRs and XOR logic without increasing the pipeline depth, both designs produce one perturbation result every $T_c$ cycles and therefore achieve identical throughput. Within the IPZO architecture, the perturbation MVM ($x\cdot\epsilon z_i$) computed by the PGU follows the same computational structure as the weight MVM ($x\cdot\theta$) in the IMC array. Each perturbation element contributes one multiplication and one addition, amounting to $2mn$ operations for an $m \times n$ PGU array. The throughput is thus given by
\begin{equation}
\text{Throughput} = \frac{2mn \cdot f}{T_c},
\end{equation}
where $f=1$ GHz and $T_c=c$ cycles. Increasing $c$ serializes more accumulation operations onto a smaller CLU array, raising $T_c$ and lowering the throughput from 4.10~TOPS at $c=1$ to 2.05, 1.02 and 0.51~TOPS at $c=2$, $4$, and $8$. Together with the area scaling at larger $c$, these results illustrate a direct tradeoff between hardware cost and throughput, in which the reduced throughput incurs no system-level penalty as long as the PGU sustains the throughput of the IMC array.

\subsubsection{Energy Consumption}

\begin{figure}[t]
  \centering
  \includegraphics[width=1\linewidth, trim={0.2cm 0.2cm 0.2cm 0.2cm}, clip]{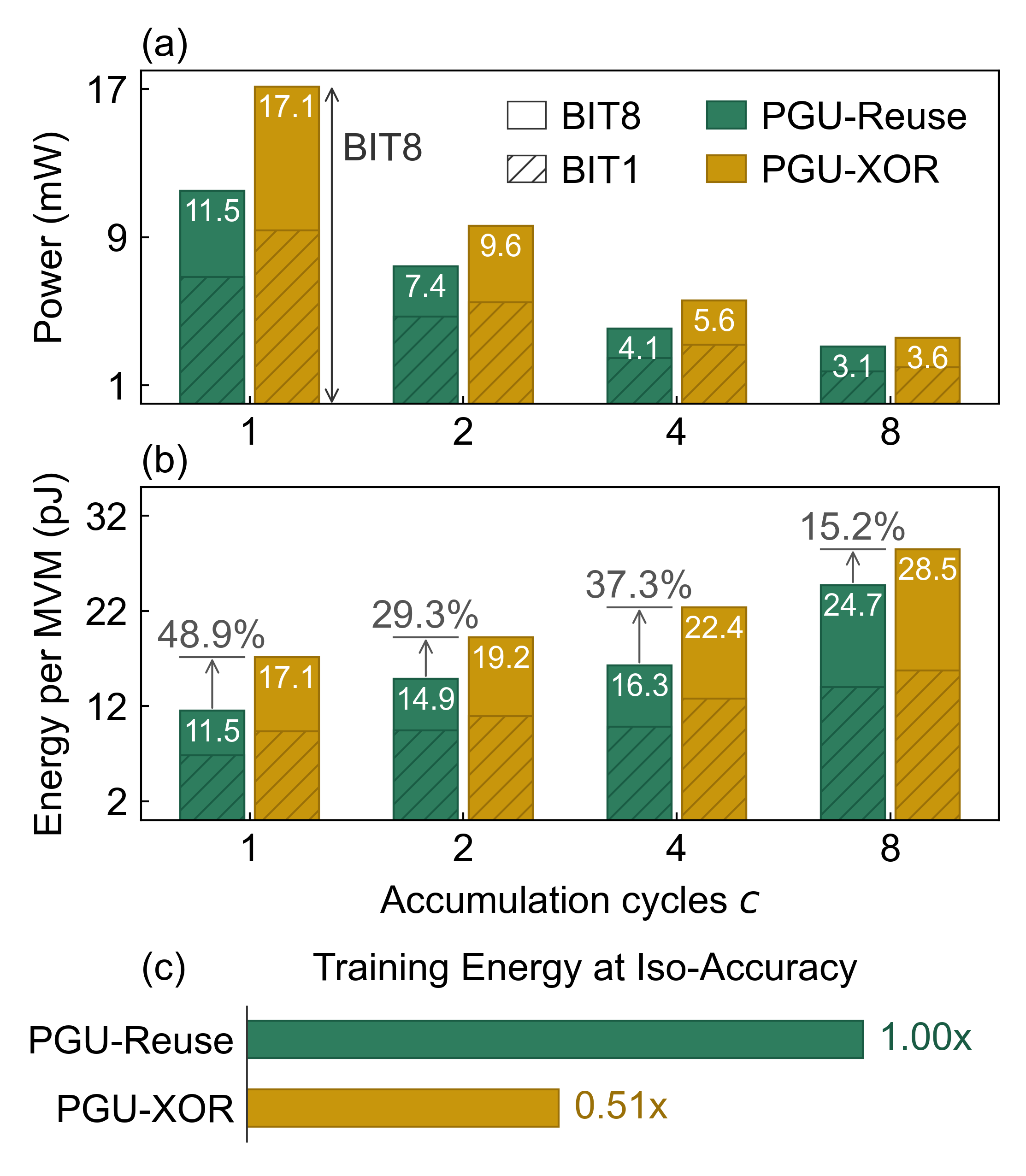}
  \caption{Post-layout (a) power and (b) energy per MVM of PGU-Reuse and PGU-XOR under different accumulation cycles $c$ in 16-nm CMOS technology, evaluated for 8-bit signed (BIT8, open bars) and 1-bit Rademacher (BIT1, hatched bars) perturbations. One MVM denotes one spike–perturbation product $x \cdot z_i$, where $x$ is the spike vector and $z_i$ is the perturbation matrix. (c) Energy cost of PGU-Reuse and PGU-XOR at iso-accuracy on CIFAR-10, normalized to PGU-Reuse.}
  \label{fig:power_energy}
\end{figure}

\begin{figure}[t]
  \centering
  \includegraphics[width=1\linewidth, trim={0.2cm 0.2cm 0.2cm 0.2cm}, clip]{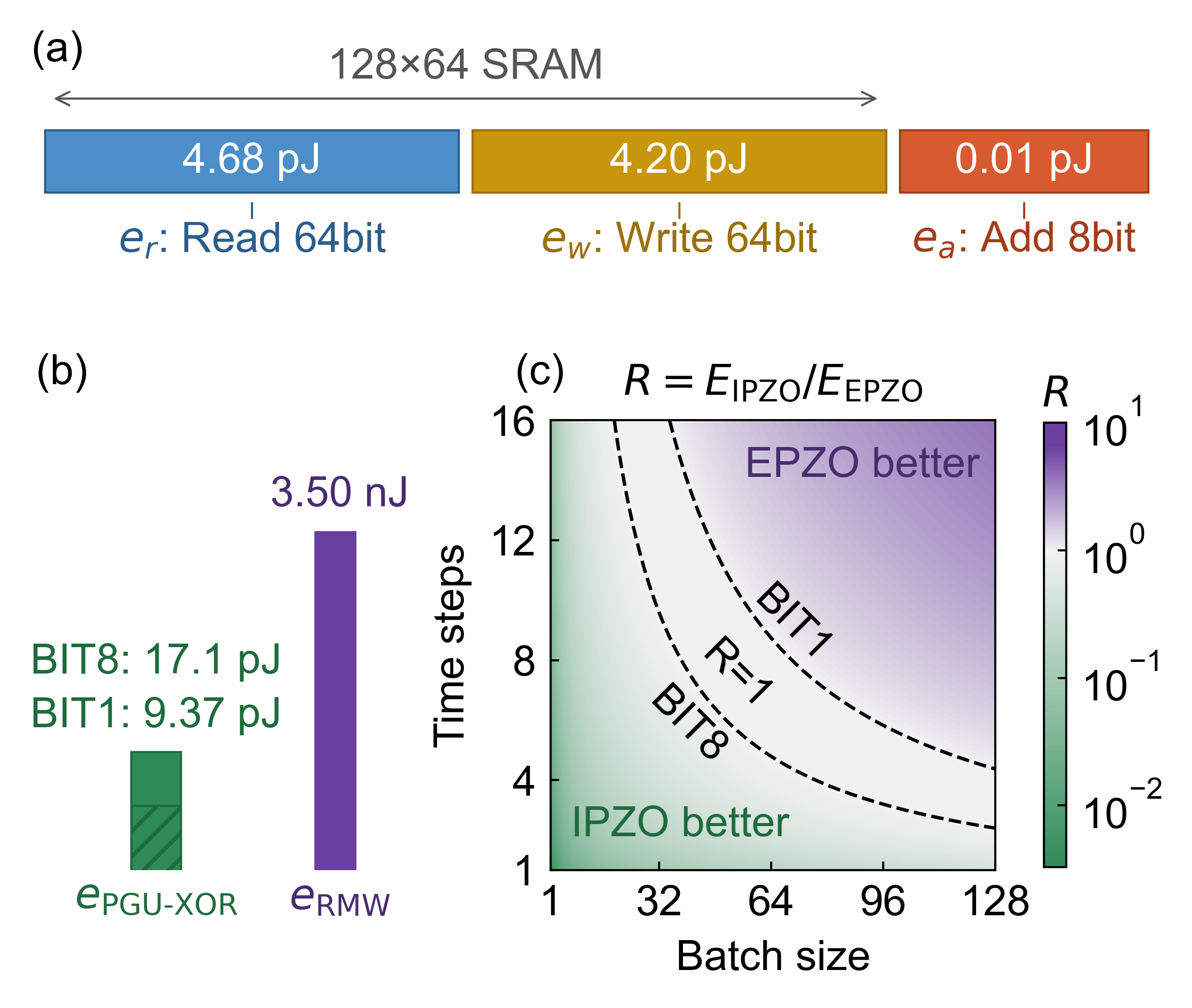}
  \caption{(a) Operation energies of a 128$\times$64 SRAM in TSMC 16-nm technology: 64-bit read $e_r$, 64-bit write $e_w$, and 8-bit addition $e_a$. (b) Energy comparison between IPZO (PGU-XOR, per MVM) obtained from post-layout simulation, and EPZO (RMW, per batch) estimated from (a). (c) Energy ratio $R = E_\mathrm{IPZO}/E_\mathrm{EPZO}$ as a function of batch size and number of time steps.}
  \label{fig:energy}
\end{figure}

Fig.~\ref{fig:power_energy} presents the post-layout power and energy of PGU-Reuse and PGU-XOR under different $c$, evaluated in both BIT8 and BIT1 perturbation modes. The energy is reported per MVM, denoting one computation between the spike vector $x$ and the perturbation matrix $z_i \in \mathbb{R}^{128\times16}$. As shown in Fig.~\ref{fig:power_energy}a, the power of both designs decreases monotonically with increasing $c$ owing to the reduced CLU array size, falling by over $3\times$ from $c=1$ to $c=8$, while BIT1 consumes about 55\%--60\% of the BIT8 power at every configuration. In contrast, the energy per MVM increases with $c$ as shown in Fig.~\ref{fig:power_energy}b because the reduced power is outweighed by the longer cycle interval $T_c$. In BIT8 mode, the energy rises from $11.5$ to $24.7$~pJ for PGU-Reuse and from $17.1$ to $28.5$~pJ for PGU-XOR. The resulting energy overhead of PGU-XOR over PGU-Reuse therefore narrows from $48.9$\% at $c=1$ to 15.2\% at $c=8$, falling below the corresponding area overhead of $40.3$\%--$46.0$\% for $c \ge 2$. This discrepancy arises because the Q-LFSRs occupy additional area but exhibit low switching activity during each optimization iteration, contributing much less dynamic energy than their physical footprint would suggest. Consequently, increasing $c$ reduces area and power at a proportional cost in throughput and energy per MVM, leaving $c$ bounded by the IMC pipeline latency and the available energy and area budget. 

The per-MVM energy advantage of PGU-Reuse, however, does not translate directly into lower energy during training. Fig.~\ref{fig:power_energy}c further compares the energy required by both PGU-Reuse and PGU-XOR to reach the same target accuracy of $66.85$\% for Spikingformer fine-tuning on CIFAR-10. Because the correlated perturbations of PGU-Reuse confine the update to a low-dimensional subspace, it requires $247$ epochs to reach this accuracy, whereas PGU-XOR converges within $84$ epochs. Despite its higher per-MVM energy, PGU-XOR therefore reduces the total perturbation energy to $0.51\times$ that of PGU-Reuse, demonstrating that the modest hardware cost of perturbation decorrelation can be offset by improved training convergence and ultimately yields a more favorable hardware--algorithm trade-off.

The energy advantage of the proposed IPZO architecture over the conventional EPZO architecture stems from the pronounced energy disparity among the primitive operations involved in weight perturbation. Fig.~\ref{fig:energy}a characterizes the energy of these operations using a $128 \times 64$ SRAM implemented in TSMC 16-nm technology. A 64-bit read ($e_r$) and a 64-bit write ($e_w$) consume $4.68$~pJ and $4.20$~pJ, respectively, whereas an 8-bit addition ($e_a$) consumes only $0.01$~pJ. This disparity of more than two orders of magnitude reveals that memory accesses, rather than arithmetic operations, dominate the energy cost of weight perturbation. IPZO exploits this gap by replacing costly weight reads and writes with low-cost additions in the accumulation domain.

Based on these primitive-operation energies, Fig.~\ref{fig:energy}b compares the estimated energy of EPZO with the post-layout energy of IPZO using PGU-XOR at $c = 1$. In EPZO, weight perturbation is performed through an RMW operation, where the weights and perturbations are read from the memory array, added element-wise, and written back. For 8-bit weights, an $m \times n$ weight matrix contains $8mn$ bits and is accessed in 64-bit words, requiring $mn/8$ read accesses each for the weights and perturbations, $mn/8$ write accesses, and $mn$ additions. The energy of one RMW operation is given by
\begin{equation}
e_{\text{RMW}} = mn\left(\frac{2e_r + e_w}{8} + e_a\right),
\end{equation}
where the factor of two on $e_r$ accounts for reading both weights and perturbations. For the $128 \times 16$ weight array, this yields $e_{\text{RMW}} = 3.50$~nJ. Each ZO optimization iteration requires three RMW operations for $q=1$, two for the positive and negative forward passes and one for the weight restoration, resulting in a fixed energy cost of $E_{\text{EPZO}} = 3\,e_{\text{RMW}} = 10.5$~nJ per optimization iteration. In contrast, the post-layout energy of PGU-XOR is $e_{\text{PGU-XOR}}=17.1$~pJ in BIT8 mode and $9.37$~pJ in BIT1 mode per input per time step. Accounting for the positive and negative evaluations, the corresponding IPZO energy scales with $2BT$.

The different scaling behaviors of the IPZO and EPZO determine the operating regime in which each architecture is more energy-efficient. To quantify this trade-off, Fig.~\ref{fig:energy}c plots the energy ratio 
\begin{equation} 
R=\frac{E_{\text{IPZO}}}{E_{\text{EPZO}}} 
=\frac{2e_{\text{PGU-XOR}}\cdot BT}{3e_{\text{RMW}}}, 
\end{equation} 
as a function of $B$ and $T$. Since $R$ scales linearly with $BT$, the break-even condition $R=1$ occurs at $BT\approx307$ in BIT8 mode and $BT\approx560$ in BIT1 mode, as indicated by the two dashed contours. Below the respective contours, IPZO is more energy-efficient, whereas above them, its accumulated per-input perturbation cost eventually exceeds the fixed RMW cost of EPZO. At $B=64$ and $T=4$, as adopted for Spikingformer fine-tuning, IPZO consumes only $0.83\times$ the energy of EPZO in BIT8 mode and $0.46\times$ in BIT1 mode. The energy advantage of IPZO becomes increasingly pronounced at smaller $BT$, which is particularly relevant to real-time on-chip learning on resource-constrained edge devices. These results demonstrate that injecting perturbations in the accumulation domain, rather than explicitly modifying stored weights, preserves the weight-stationary execution of IMC while substantially reducing perturbation energy across practical training configurations.

\section{Conclusion}
This work presented IPZO, which is a hardware-algorithm co-designed architecture for efficient ZO-based on-chip learning in IMC-based spiking transformers. IPZO relocates random perturbations from the stored weight domain to the accumulation domain, eliminating the repeated RMW operations of explicit weight perturbation while preserving weight-stationary IMC execution. Within IPZO, the event-triggered PGU-XOR generates perturbations only for spike-activated rows, decoupling the RNG array size from the weight matrix dimensions, while address-driven XOR recombination eliminates the spatial correlations introduced by the resulting RNG reuse. PGU-XOR achieves near-full-rank exploration of the parameter space and optimization performance comparable to software RNGs across both image classification and language modeling. Post-layout implementation in TSMC 16-nm CMOS shows that, despite the additional hardware for perturbation decorrelation, PGU-XOR reduces the total perturbation energy to $0.51\times$ that of PGU-Reuse through faster convergence. At the architecture level, IPZO with PGU-XOR reduces the perturbation energy to $0.46$--$0.83\times$ that of EPZO at $B=64$ and $T=4$. These results demonstrate that combining accumulation-domain perturbation with sparsity-driven, decorrelated perturbation generation provides an efficient hardware approach to ZO-based on-chip learning for IMC-based SNN accelerators.

% \bibliographystyle{IEEEtran}
% \bibliography{references}

\bibliographystyle{IEEEtran}
\bibliography{references}

@article{vaswani2017attention,
  title={Attention is all you need},
  author={Vaswani, Ashish and Shazeer, Noam and Parmar, Niki and Uszkoreit, Jakob and Jones, Llion and Gomez, Aidan N and Kaiser, {\L}ukasz and Polosukhin, Illia},
  journal={Advances in neural information processing systems},
  volume={30},
  year={2017}
}

@inproceedings{wolf2020transformers,
  title={Transformers: State-of-the-art natural language processing},
  author={Wolf, Thomas and Debut, Lysandre and Sanh, Victor and Chaumond, Julien and Delangue, Clement and Moi, Anthony and Cistac, Pierric and Rault, Tim and Louf, R{\'e}mi and Funtowicz, Morgan and others},
  booktitle={Proceedings of the 2020 conference on empirical methods in natural language processing: system demonstrations},
  pages={38--45},
  year={2020}
}

@article{wu2020visual,
  title={Visual transformers: Token-based image representation and processing for computer vision},
  author={Wu, Bichen and Xu, Chenfeng and Dai, Xiaoliang and Wan, Alvin and Zhang, Peizhao and Yan, Zhicheng and Tomizuka, Masayoshi and Gonzalez, Joseph and Keutzer, Kurt and Vajda, Peter},
  journal={arXiv preprint arXiv:2006.03677},
  year={2020}
}

@article{han2022survey,
  title={A survey on vision transformer},
  author={Han, Kai and Wang, Yunhe and Chen, Hanting and Chen, Xinghao and Guo, Jianyuan and Liu, Zhenhua and Tang, Yehui and Xiao, An and Xu, Chunjing and Xu, Yixing and others},
  journal={IEEE transactions on pattern analysis and machine intelligence},
  volume={45},
  number={1},
  pages={87--110},
  year={2022},
  publisher={IEEE}
}

@article{wolters2024memory,
  title={Memory is all you need: An overview of compute-in-memory architectures for accelerating large language model inference},
  author={Wolters, Christopher and Yang, Xiaoxuan and Schlichtmann, Ulf and Suzumura, Toyotaro},
  journal={arXiv preprint arXiv:2406.08413},
  year={2024}
}

@article{roy2019towards,
  title={Towards spike-based machine intelligence with neuromorphic computing},
  author={Roy, Kaushik and Jaiswal, Akhilesh and Panda, Priyadarshini},
  journal={Nature},
  volume={575},
  number={7784},
  pages={607--617},
  year={2019},
  publisher={Nature Publishing Group UK London}
}

@article{davies2018loihi,
  title={Loihi: A neuromorphic manycore processor with on-chip learning},
  author={Davies, Mike and Srinivasa, Narayan and Lin, Tsung-Han and Chinya, Gautham and Cao, Yongqiang and Choday, Sri Harsha and Dimou, Georgios and Joshi, Prasad and Imam, Nabil and Jain, Shweta and others},
  journal={Ieee micro},
  volume={38},
  number={1},
  pages={82--99},
  year={2018},
  publisher={IEEE}
}

@article{dampfhoffer2023backpropagation,
  title={Backpropagation-based learning techniques for deep spiking neural networks: A survey},
  author={Dampfhoffer, Manon and Mesquida, Thomas and Valentian, Alexandre and Anghel, Lorena},
  journal={IEEE Transactions on Neural Networks and Learning Systems},
  volume={35},
  number={9},
  pages={11906--11921},
  year={2023},
  publisher={IEEE}
}

@inproceedings{meng2023towards,
  title={Towards memory-and time-efficient backpropagation for training spiking neural networks},
  author={Meng, Qingyan and Xiao, Mingqing and Yan, Shen and Wang, Yisen and Lin, Zhouchen and Luo, Zhi-Quan},
  booktitle={2023 IEEE/CVF International Conference on Computer Vision (ICCV)},
  pages={6143--6153},
  year={2023},
  organization={IEEE}
}

@inproceedings{zhang2024memory,
  title={Memory-efficient reversible spiking neural networks},
  author={Zhang, Hong and Zhang, Yu},
  booktitle={Proceedings of the AAAI conference on artificial intelligence},
  volume={38},
  number={15},
  pages={16759--16767},
  year={2024}
}

@inproceedings{qinzeroth,
  title={Zeroth-Order Forward-Only SNN Training Inspiring Neuromorphic On-Chip Learning},
  author={Qin, Mingyue and Yin, Shuyu and Guo, Qinghai and Liu, Peilin and Huang, Xiaolin and Wen, Fei},
  booktitle={Forty-third International Conference on Machine Learning}
}

@inproceedings{liu2018zeroth,
  title={Zeroth-order stochastic projected gradient descent for nonconvex optimization},
  author={Liu, Sijia and Li, Xingguo and Chen, Pin-Yu and Haupt, Jarvis and Amini, Lisa},
  booktitle={2018 IEEE Global Conference on Signal and Information Processing (GlobalSIP)},
  pages={1179--1183},
  year={2018},
  organization={IEEE}
}

@article{liu2020primer,
  title={A primer on zeroth-order optimization in signal processing and machine learning: Principals, recent advances, and applications},
  author={Liu, Sijia and Chen, Pin-Yu and Kailkhura, Bhavya and Zhang, Gaoyuan and Hero III, Alfred O and Varshney, Pramod K},
  journal={IEEE Signal Processing Magazine},
  volume={37},
  number={5},
  pages={43--54},
  year={2020},
  publisher={IEEE}
}

@inproceedings{chen2025analog,
  title={Analog Multilevel eDRAM-RRAM CIM for Zeroth-Order Fine-tuning of LLMs},
  author={Chen, Mufeng and Zheng, Luqi and Lin, Jian-Yu and Ye, Peide D and Li, Haitong},
  booktitle={2025 IEEE International Memory Workshop (IMW)},
  pages={1--4},
  year={2025},
  organization={IEEE}
}

@inproceedings{wang2025noisezo,
  title={NoiseZO: RRAM Noise-Driven Zeroth-Order Optimization for Efficient Forward-Only Training},
  author={Wang, Shuqi and Liu, Zhengwu and Ding, Chenchen and Zhang, Chen and Wu, Taiqiang and Zhou, Jiajun and Wong, Ngai},
  booktitle={2025 62nd ACM/IEEE Design Automation Conference (DAC)},
  pages={1--7},
  year={2025},
  organization={IEEE}
}

@article{shen2024efficient,
  title={On efficient training of large-scale deep learning models},
  author={Shen, Li and Sun, Yan and Yu, Zhiyuan and Ding, Liang and Tian, Xinmei and Tao, Dacheng},
  journal={ACM Computing Surveys},
  volume={57},
  number={3},
  pages={1--36},
  year={2024},
  publisher={ACM New York, NY}
}

@article{sinangil20207,
  title={A 7-nm compute-in-memory SRAM macro supporting multi-bit input, weight and output and achieving 351 TOPS/W and 372.4 GOPS},
  author={Sinangil, Mahmut E and Erbagci, Burak and Naous, Rawan and Akarvardar, Kerem and Sun, Dar and Khwa, Win-San and Liao, Hung-Jen and Wang, Yih and Chang, Jonathan},
  journal={IEEE Journal of Solid-State Circuits},
  volume={56},
  number={1},
  pages={188--198},
  year={2020},
  publisher={IEEE}
}

@article{jiang2020c3sram,
  title={C3SRAM: An in-memory-computing SRAM macro based on robust capacitive coupling computing mechanism},
  author={Jiang, Zhewei and Yin, Shihui and Seo, Jae-Sun and Seok, Mingoo},
  journal={IEEE Journal of Solid-State Circuits},
  volume={55},
  number={7},
  pages={1888--1897},
  year={2020},
  publisher={IEEE}
}

@inproceedings{tan2025perturbation,
  title={Perturbation-efficient zeroth-order optimization for hardware-friendly on-device training},
  author={Tan, Qitao and Chang, Sung-En and Xia, Rui and Ji, Huidong and Yang, Chence and Zhang, Ci and Liu, Jun and Zhan, Zheng and Fang, Zhenman and Zou, Zhuo and others},
  booktitle={2025 IEEE/ACM International Conference On Computer Aided Design (ICCAD)},
  pages={1--9},
  year={2025},
  organization={IEEE}
}

@inproceedings{wu202499,
  title={A 99.2 TOPS/W Transformer Learning Processor with Approximated Attention Score Gradient Computation and Ternary Vector-based Speculation},
  author={Wu, Ping-Sheng and Lin, Yu-Cheng and Yang, Chia-Hsiang},
  booktitle={2024 IEEE Symposium on VLSI Technology and Circuits (VLSI Technology and Circuits)},
  pages={1--2},
  year={2024},
  organization={IEEE}
}

@inproceedings{suwandi2026breaking,
  title={Breaking the Curse of Dimensionality in Gaussian Process Training with zeroth-order Adaptive Perturbation},
  author={Suwandi, Richard Cornelius and Yin, Feng and Chang, Tsung-Hui},
  booktitle={ICASSP 2026-2026 IEEE International Conference on Acoustics, Speech and Signal Processing (ICASSP)},
  pages={22497--22501},
  year={2026},
  organization={IEEE}
}

@inproceedings{yeh202616nm,
  title={A 16nm, 1Mb, 1-to-8b-Configurable 444.21 TOPS/W Fully Digital SRAM Compute-in-Memory Macro for Hybrid SNN-CNN Edge Computing},
  author={Yeh, Yao-Kai and Su, Jian-Wei and Hsu, Ting-Hao and Tseng, Mai and Tien, Jen-Chun and Chen, Ko-Chi and Yue, Chih-Yen and Lin, Yu-En and Hu, Yu-Jia and Hsieh, Le-Jung and others},
  booktitle={2026 IEEE International Solid-State Circuits Conference (ISSCC)},
  volume={69},
  pages={526--528},
  year={2026},
  organization={IEEE}
}

@article{malladi2023fine,
  title={Fine-tuning language models with just forward passes},
  author={Malladi, Sadhika and Gao, Tianyu and Nichani, Eshaan and Damian, Alex and Lee, Jason D and Chen, Danqi and Arora, Sanjeev},
  journal={Advances in Neural Information Processing Systems},
  volume={36},
  pages={53038--53075},
  year={2023}
}

@inproceedings{wang202422nm,
  title={A 22nm 54.94 TFLOPS/W transformer fine-tuning processor with exponent-stationary re-computing, aggressive linear fitting, and logarithmic domain multiplicating},
  author={Wang, Yang and Yang, Xiaolong and Qin, Yubin and Zhao, Zhiren and Guo, Ruiqi and Yue, Zhiheng and Han, Huiming and Wei, Shaojun and Hu, Yang and Yin, Shouyi},
  booktitle={2024 IEEE Symposium on VLSI Technology and Circuits (VLSI Technology and Circuits)},
  pages={1--2},
  year={2024},
  organization={IEEE}
}

@article{larson2019derivative,
  title={Derivative-free optimization methods},
  author={Larson, Jeffrey and Menickelly, Matt and Wild, Stefan M},
  journal={Acta Numerica},
  volume={28},
  pages={287--404},
  year={2019},
  publisher={Cambridge University Press}
}

@book{han2023chip,
  title={On-Chip Training NPU-Algorithm, Architecture and SoC Design},
  author={Han, Donghyeon and Yoo, Hoi-Jun},
  year={2023},
  publisher={Springer}
}

@inproceedings{nag2023vita,
  title={ViTA: A vision transformer inference accelerator for edge applications},
  author={Nag, Shashank and Datta, Gourav and Kundu, Souvik and Chandrachoodan, Nitin and Beerel, Peter A},
  booktitle={2023 IEEE International Symposium on Circuits and Systems (ISCAS)},
  pages={1--5},
  year={2023},
  organization={IEEE}
}

@article{mun2026asap,
  title={ASAP: A 28-nm Transformer Training Accelerator With Alternating Sparsity and Asymmetrical Microscaling Precision},
  author={Mun, HanGyeol and Meng, Jian and Hu, Xiaofeng and Liao, Yuan and Chen, Chun-Ting and Seo, Jae-sun},
  journal={IEEE Journal of Solid-State Circuits},
  year={2026},
  publisher={IEEE}
}

@article{keller202395,
  title={A 95.6-TOPS/W deep learning inference accelerator with per-vector scaled 4-bit quantization in 5 nm},
  author={Keller, Ben and Venkatesan, Rangharajan and Dai, Steve and Tell, Stephen G and Zimmer, Brian and Sakr, Charbel and Dally, William J and Gray, C Thomas and Khailany, Brucek},
  journal={IEEE Journal of Solid-State Circuits},
  volume={58},
  number={4},
  pages={1129--1141},
  year={2023},
  publisher={IEEE}
}

@article{song2025xpikeformer,
  title={Xpikeformer: Hybrid analog-digital hardware acceleration for spiking transformers},
  author={Song, Zihang and Katti, Prabodh and Simeone, Osvaldo and Rajendran, Bipin},
  journal={IEEE Transactions on Very Large Scale Integration (VLSI) Systems},
  volume={33},
  number={6},
  pages={1596--1609},
  year={2025},
  publisher={IEEE}
}

@article{das2025aster,
  title={ASTER: Attention-based Spiking Transformer Engine for Event-driven Reasoning},
  author={Das, Tamoghno and Vu, Khanh Phan and Chen, Hanning and Oh, Hyunwoo and Imani, Mohsen},
  journal={arXiv preprint arXiv:2511.06770},
  year={2025}
}

@inproceedings{jiang2025sparta,
  title={SPARTA: Spike-Aware Token Skipping Co-Optimization with Heterogeneous ReRAM-CIM Architecture for Spiking Transformer Acceleration},
  author={Jiang, Pinfeng and Wang, Letian and Fang, Yilong and Wang, Yi and Zhu, Mingde and Miao, Xiangshui and Wang, Xingsheng},
  booktitle={2025 IEEE/ACM International Conference On Computer Aided Design (ICCAD)},
  pages={1--9},
  year={2025},
  organization={IEEE}
}

@article{zhang2024revisiting,
  title={Revisiting zeroth-order optimization for memory-efficient llm fine-tuning: A benchmark},
  author={Zhang, Yihua and Li, Pingzhi and Hong, Junyuan and Li, Jiaxiang and Zhang, Yimeng and Zheng, Wenqing and Chen, Pin-Yu and Lee, Jason D and Yin, Wotao and Hong, Mingyi and others},
  journal={arXiv preprint arXiv:2402.11592},
  year={2024}
}

@article{sugiura2025elasticzo,
  title={Elasticzo: A memory-efficient on-device learning with combined zeroth-and first-order optimization},
  author={Sugiura, Keisuke and Matsutani, Hiroki},
  journal={arXiv preprint arXiv:2501.04287},
  year={2025}
}

@inproceedings{gu2021efficient,
  title={Efficient on-chip learning for optical neural networks through power-aware sparse zeroth-order optimization},
  author={Gu, Jiaqi and Feng, Chenghao and Zhao, Zheng and Ying, Zhoufeng and Chen, Ray T and Pan, David Z},
  booktitle={Proceedings of the AAAI conference on artificial intelligence},
  volume={35},
  number={9},
  pages={7583--7591},
  year={2021}
}

@inproceedings{chen2024deepzero,
  title={Deepzero: Scaling up zeroth-order optimization for deep model training},
  author={Chen, Aochuan and Zhang, Yimeng and Jia, Jinghan and Diffenderfer, James and Parasyris, Konstantinos and Liu, Jiancheng and Zhang, Yihua and Zhang, Zheng and Kailkhura, Bhavya and Liu, Sijia},
  booktitle={International Conference on Learning Representations},
  volume={2024},
  pages={50185--50206},
  year={2024}
}

@inproceedings{zhou2026spikingformer,
  title={Spikingformer: A key foundation model for spiking neural networks},
  author={Zhou, Chenlin and Yu, Liutao and Zhou, Zhaokun and Zhang, Han and Wang, Jiaqi and Zhou, Huihui and Ma, Zhengyu and Tian, Yonghong},
  booktitle={Proceedings of the AAAI Conference on Artificial Intelligence},
  volume={40},
  number={3},
  pages={2236--2244},
  year={2026}
}

@article{
zhu2024spikegpt,
title={Spike{GPT}: Generative Pre-trained Language Model with Spiking Neural Networks},
author={Rui-Jie Zhu and Qihang Zhao and Guoqi Li and Jason Eshraghian},
journal={Transactions on Machine Learning Research},
issn={2835-8856},
year={2024},
url={https://openreview.net/forum?id=gcf1anBL9e},
note={}
}

@article{sharma2022reconfigurable,
  title={A reconfigurable 16Kb AND8T SRAM macro with improved linearity for multibit compute-in memory of artificial intelligence edge devices},
  author={Sharma, Vishal and Kim, Ju-Eon and Kim, Hyunjoon and Lu, Lu and Kim, Tony Tae-Hyoung},
  journal={IEEE Journal on Emerging and Selected Topics in Circuits and Systems},
  volume={12},
  number={2},
  pages={522--535},
  year={2022},
  publisher={IEEE}
}

@article{stillmaker2017scaling,
  title={Scaling equations for the accurate prediction of CMOS device performance from 180 nm to 7 nm},
  author={Stillmaker, Aaron and Baas, Bevan},
  journal={Integration},
  volume={58},
  pages={74--81},
  year={2017},
  publisher={Elsevier}
}

@inproceedings{chich202189,
  title={An 89 TOPS/W and 16.3 TOPS/MM 2 all-digital SRAM-based full-precision compute-in memory macro in 22nm for machine-learning edge applications},
  author={Chich, YD and others},
  booktitle={ISSCC},
  pages={252--253},
  year={2021}
}

@article{fournier2023practical,
  title={A practical survey on faster and lighter transformers},
  author={Fournier, Quentin and Caron, Ga{\'e}tan Marceau and Aloise, Daniel},
  journal={ACM Computing Surveys},
  volume={55},
  number={14s},
  pages={1--40},
  year={2023},
  publisher={ACM New York, NY}
}

@INPROCEEDINGS{10595893,
  author={Song, Zihang and Katti, Prabodh and Simeone, Osvaldo and Rajendran, Bipin},
  booktitle={2024 IEEE 6th International Conference on AI Circuits and Systems (AICAS)}, 
  title={Stochastic Spiking Attention: Accelerating Attention with Stochastic Computing in Spiking Networks}, 
  year={2024},
  volume={},
  number={},
  pages={31-35},
  doi={10.1109/AICAS59952.2024.10595893}}

@article{cessac2011discrete,
  title = {A Discrete Time Neural Network Model with Spiking Neurons {{II}}. {{Dynamics}} with Noise},
  author = {Cessac, B.},
  year = {2011},
  journal = {Journal of Mathematical Biology},
  volume = {62},
  number = {6},
  eprint = {1002.3275},
  primaryclass = {physics, q-bio},
  pages = {863--900},
  doi = {10.1007/s00285-010-0358-4},
  urldate = {2023-11-29},
  archiveprefix = {arxiv}
}

@article{jhang2021challenges,
  title={{Challenges and trends of SRAM-based computing-in-memory for AI edge devices}},
  author={Jhang, Chuan-Jia and Xue, Cheng-Xin and Hung, Je-Min and Chang, Fu-Chun and Chang, Meng-Fan},
  journal={IEEE Transactions on Circuits and Systems I: Regular Papers},
  volume={68},
  number={5},
  pages={1773--1786},
  year={2021},
  publisher={IEEE}
}

@article{kneip2021impact,
  title={{Impact of analog non-idealities on the design space of 6T-SRAM current-domain dot-product operators for in-memory computing}},
  author={Kneip, Adrian and Bol, David},
  journal={IEEE Transactions on Circuits and Systems I: Regular Papers},
  volume={68},
  number={5},
  pages={1931--1944},
  year={2021},
  publisher={IEEE}
}

@article{aguirre2024hardware,
  title={Hardware implementation of memristor-based artificial neural networks},
  author={Aguirre, Fernando and Sebastian, Abu and Le Gallo, Manuel and Song, Wenhao and Wang, Tong and Yang, J Joshua and Lu, Wei and Chang, Meng-Fan and Ielmini, Daniele and Yang, Yuchao and others},
  journal={Nature communications},
  volume={15},
  number={1},
  pages={1974},
  year={2024},
  publisher={Nature Publishing Group UK London}
}

@inproceedings{khaddam2021hermes,
  title={{HERMES} Core--A 14nm {CMOS} and PCM-based In-Memory Compute Core using an array of 300ps/{LSB} Linearized {CCO}-based {ADCs} and local digital processing},
  author={Khaddam-Aljameh, Riduan and Stanisavljevic, Milos and Mas, J Fornt and Karunaratne, Geethan and Braendli, Matthias and Liu, Femg and Singh, Abhairaj and M{\"u}ller, Silvia M and Egger, Urs and Petropoulos and others},
  booktitle={2021 Symposium on VLSI Circuits},
  year={2021},
  organization={IEEE}
}

@article{verma2023neuromorphic,
  title={Neuromorphic Accelerator for Spiking Neural Network Using {SOT-MRAM} Crossbar Array},
  author={Verma, Gaurav and Nisar, Arshid and Dhull, Seema and Kaushik, Brajesh Kumar},
  journal={IEEE Transactions on Electron Devices},
  year={2023},
  publisher={IEEE}
}

\vfill

\end{document}